\documentclass[prd,aps,twocolumn,floats,floatfix,nofootinbib]{revtex4-1}

\usepackage[utf8]{inputenc}
\usepackage{amssymb,bm}
\usepackage{graphicx}
\usepackage{dcolumn}
\usepackage{bm}
\usepackage{graphics}
\usepackage{slashed}
\usepackage{enumerate}
\usepackage{amsmath}
\usepackage{url}
\usepackage[usenames,dvipsnames,svgnames,table]{xcolor}
\usepackage{color}
\usepackage{xcolor}
\usepackage{ulem}
\usepackage{comment}

\definecolor{color1}{RGB}{191, 0, 255}
\def\CP#1{{\textcolor{blue}{CP: #1}}}

\def\FC#1{{\textcolor{red}{FC: #1}}}

\begin{document}

\title{The role of turbulence and winding in the development of large-scale, strong magnetic fields in long-lived remnants of binary neutron star mergers}

\author{
	Ricard Aguilera-Miret$^{1,3}$,
	Carlos Palenzuela$^{2,3,4}$	
	Federico Carrasco$^{5,3}$,
	Daniele Vigan\`o$^{6,4,3}$,
}

\affiliation{${^1}$University of Hamburg, Hamburger Sternwarte, Gojenbergsweg 112, 21029, Hamburg, Germany}
\affiliation{${^2}$Departament  de  F\'{\i}sica,  Universitat  de  les  Illes  Balears,  Palma  de  Mallorca, E-07122,  Spain}
\affiliation{$^3$Institute of Applied Computing \& Community Code (IAC3),  Universitat  de  les  Illes  Balears,  Palma  de  Mallorca, E-07122,  Spain}
\affiliation{$^4$Institut d'Estudis Espacials de Catalunya (IEEC), 08034 Barcelona, Spain}
\affiliation{$^5$Instituto de F\'isica Enrique Gaviola, CONICET-UNC, 5000 C\'ordoba, Argentina}
\affiliation{$^{6}$Institute of Space Sciences (IEEC-CSIC), E-08193 Barcelona, Spain}

\begin{abstract}
We perform a long and accurate Large-Eddy Simulation of a binary neutron star merger, following the newly formed remnant up to $110$ milliseconds. The combination of high-order schemes, high-resolution and the gradient subgrid-scale model allow us to have among the highest effective resolutions ever achieved.
Our results show that, although the magnetic fields are strongly amplified by the Kelvin-Helmholtz instability, they are coherent only over very short spatial scales until $t \gtrsim 30$~ms. Around that time, magnetic winding becomes more efficient leading to a linear growth of the toroidal component and slowly ordering the field to more axisymmetric, large scales. The poloidal component only starts to grow at small scales at much later times $t \gtrsim 90$~ms, in a way compatible with the magneto-rotational instability. No strong large-scale poloidal field or jet is produced in the timescales spanned by our simulation, although there is an helicoidal structure gradually developing at late times. We highlight that soon after the merger the topology is always strongly dominated by toroidal structures, with a complex distribution in the meridional plane and highly turbulent perturbations. Thus, starting with strong purely dipolar fields before the merger is largely inconsistent with the outcomes of a realistic evolution.
Finally, we confirm the universality of the evolved topology, even when starting with very different magnetic fields confined to the outermost layers of each neutron star.
\end{abstract}

\maketitle

\section{Introduction}

The tremendous scientific potential of binary neutron star (BNS) mergers, detected simultaneously through gravitational waves (GW) and electromagnetic (EM) counterparts, was confirmed with the event GW170817~\cite{LVC-BNS,LVC-MMA}, of which the current Ligo/Virgo/Kagra O4-run is hoped to bring similar detections. Among other astrophysical information, this event provided the most compelling evidence that BNS mergers can produce powerful jets and SGRBs~\cite{LVC-GRB,goldstein2017,savchenko2017,Troja2017,Margutti2017,Hallinan2017,Alexander2017,Mooley2018a,Lazzati2018,Lyman2018,Alexander2018,Mooley2018b,Ghirlanda2019}, as well as copious amounts of heavy r-process elements (e.g.,~\cite{Arcavi2017,Coulter2017,Pian2017,Smartt2017,Kasen2017,Metzger2019LRR} and references therein).
Although the binary inspiral is mostly driven by gravitational wave emission, several non-linear magnetohydrodynamic (MHD) processes and instabilities affects the post-merger dynamics. For this reason, some key elements of the  matter and magnetic fields remnant's evolution remain still uncertain, including the actual mechanisms behind jet formation and matter ejection (e.g.,~\cite{Ciolfi2020c}).

The only way to accurately model the remnant's dynamics in a realistic manner is through numerical relativity simulations of merging BNSs. These simulations require the inclusion of fundamental ingredients such as magnetic fields, temperature and composition dependent equation of state (EoS), and neutrino radiation (see, e.g.,~\cite{Paschalidis2017,Duez2019,Shibata2019,Ciolfi2020b,Palenzuela2020} for recent reviews). Here we focus on the influence of the magnetic fields, which are known to play an important key role in the remnant's evolution and in shaping the electromagnetic counterpart signals. Strong, large-scale magnetic fields are believed to be necessary to power relativistic jets, and magnetic field stresses might contribute significantly to the secular ejecta producing part of the kilonova spectra (e.g.,~\cite{Ciolfi2020b} and references therein). However, adding a magnetic field in BNS simulations increases significantly their complexity and poses new computational challenges.

Several works have found that the magnetic field energy is amplified during the merger process up to $10^{51}$~ergs or even higher. This magnetic field amplification relies on different mechanisms: the winding effect by the large-scale differential rotation (\cite{Duez2006b}), the Kelvin-Helmholtz instability (KHI, e.g.,~\cite{price06,kiuchi15}), and the magneto-rotational instability (MRI,~\cite{balbus91,balbus98,duez2006a,siegel13,kiuchi14}). The latter two generally happen at (very) small scales which cannot be fully captured with the standard resolutions currently used in numerical simulations. The highest resolution (and computationally most expensive) simulations ever performed are far from completely resolving these instabilities along with the associated magnetohydrodynamics turbulence~\cite{kiuchi18,2023arXiv230615721K}. Different approaches have been proposed to overcome the above limitations and incorporate at least partially the effects of unresolved MHD processes.

One of the most common approaches in BNS simulations is to impose a very strong, purely poloidal, large-scale magnetic field $\gtrsim 10^{14}$G just before the merger (e.g.,~\cite{ruiz16,kiuchi18,ciolfi2019,ciolfi2020collimated,ruiz2020,mosta2020,combi2023b,Most2023,2023arXiv230615721K}). The choice of such large magnetic field intensity is qualitatively justified to compensate the inability of capturing the above-mentioned MHD instabilities when employing the more realistic values $\lesssim 10^{11}$G expected in Gyr-old NSs~\cite{tauris17}. However, this commonly chosen large-scale magnetic field is at complete contrast with the mostly turbulent and eventually toroidally dominated topology developed during the merger, regardless of the initial configuration and intensity~\cite{aguilera2020,palenzuela22,aguilera22,kiuchi18,mosta2020}. 

An interesting alternative to compensate for the lack of resolution relies on the use of Large-Eddy Simulations (LESs), a technique in which the general relativistic MHD (GRMHD) evolution equations are modified by including new terms to account for the unresolved subgrid-scale (SGS) dynamics (e.g.,~\cite{zhiyin15}). In particular, the gradient SGS model~\cite{leonard75,muller02a,grete16,grete17phd} is a sophisticated choice, conceptually similar to a numerical mathematically-informed flux reconstruction, with no a-priori physical assumptions. By including these SGS terms in the equations one can recover, at least partially, the effects induced by the unresolved SGS dynamics over the resolved scales. The combination of high-order numerical schemes, high-resolution and LES with the gradient SGS model allowed us to improve the accuracy of the magnetic amplification in a turbulent regime, both in box~\cite{carrasco19,vigano20} and in BNS merger simulations~\cite{aguilera2020,palenzuela22,aguilera22}. 
In particular, we found convergence on the (averaged) magnetic fields reached after the KHI amplification phase~\cite{palenzuela22}.
In this work, we improve and extend the BNS simulation performed in Ref.~\cite{palenzuela22}, focusing on the magnetic field evolution in long timescales after merger. Compared to previous works, we further deepen our analysis and try to pinpoint the underlying most relevant physical mechanisms at every stage.

Moreover, a recent work claimed important differences in the evolved topology if one starts with an initial magnetic field confined to the outermost layers of each NS~\cite{chabanov23} (i.e., vaguely corresponding to the NS crust). These results are in strong contrast with the universality of the final topology that we found in Ref.~\cite{aguilera22}. In particular, our simulations showed that the initial configuration of the magnetic field is quickly forgotten, as long as: (i) it is not initially set up with unrealistic values $ \gtrsim 10^{14}$G, and (ii) one can numerically resolve the small-scale instabilities that cause a turbulent magnetic field amplification in the first milliseconds after the merger. In order to clarify this controversy, we repeated their numerical experiment by comparing the KHI phase for their same two topologies.

The paper is organized as follows. The numerical setup, which includes evolution equations, numerical schemes, EoS, initial conditions and analysis quantities, is briefly summarized in~\S\ref{sec:setup}. Then, numerical results are presented and analyzed in~\S\ref{sec:results}. It follows in~\S\ref{sec:discussion} an explanation of the physical processes and instabilities occurring in the remnant, with the supporting evidences of the results and a theoretical background. Next, in~\S\ref{sec:outermost_layers} we repeat the simulations of \cite{chabanov23}, comparing the outcomes. Finally, the conclusions are drawn in \S\ref{sec:conclusions}. 

\section{Numerical setup} \label{sec:setup}

The concept and mathematical foundations behind LES with a gradient SGS approach have been extensively explained in our previous works (and references therein) in the context of classical~\cite{vigano19b} and relativistic MHD~\cite{carrasco19,vigano20,aguilera2020,palenzuela22,aguilera22}, to which we refer for details and further references. In the present work, we will perform LESs of the full GRMHD equations with the gradient SGS model, extending the simulations presented in~\cite{palenzuela22}. Initial data, evolution equations, numerical schemes and setup are almost identical as in~\cite{palenzuela22}, which we now summarize briefly for completeness.

\subsection{Evolution equations: GRMHD LES} \label{sec:equations}

The spacetime geometry is described by the Einstein equations.
The covariant field equations can be written as an hyperbolic evolution system by performing the $3+1$ decomposition (see, e.g.,~\cite{bonabook,Palenzuela2020}). Here, we use the covariant conformal Z4 formulation of the evolution equations~\cite{alic12,bezares17}. A summary of the final set of evolution equations for the spacetime fields, together with the gauge conditions setting the choice of coordinates, can be found in~\cite{palenzuela18}.
The magnetized perfect fluid describing the star follows the GRMHD equations (see, e.g., \cite{shibatabook,palenzuela15}), a set of evolution equations for the conserved variables $ \left\lbrace D, S^i , U, B^i \right\rbrace$. These conserved fields are functions of the primitive fields, namely the rest-mass density $\rho$, the specific internal energy $\epsilon$, the velocity vector $v^{i}$ and the magnetic field $B^i$. The recovery of the primitive from the conserved fields requires first a closure relation for the pressure $p$ (i.e., the EoS) and then solving a set of non-linear algebraic equations involving the Lorentz factor $W = (1-v^2)^{-1/2}$, as it is discussed in Sec.~\ref{sec:EoS}. The full set of evolution equations, including all the gradient SGS terms, can be found in~\cite{vigano20,aguilera2020}.

Each SGS term has a free parameter of order unity, which needs to be magnified to compensate for the numerical dissipation of the employed numerical scheme. Following our previous studies~\cite{carrasco19,vigano20,aguilera2020,palenzuela22,aguilera22}, and since we are mostly interested in the magnetic field dynamics, we include only the SGS term appearing on the induction equation with the pre-coefficient ${\cal C_M}= 8$, which has been shown to reproduce the magnetic field amplification more accurately (i.e., comparing with very high-resolution simulations) for our numerical schemes~\cite{vigano20,aguilera2020}. We remind the reader that these SGS terms, by construction, vanish at the continuous limit.  

\subsection{Numerical methods}

As in our previous works, and in particular~\cite{aguilera2020,palenzuela22,aguilera22}, we use the code {\sc MHDuet}, generated by the platform {\sc Simflowny} \cite{arbona13,arbona18} and based on the {\sc SAMRAI} infrastructure \cite{hornung02,gunney16}, which provides the parallelization and the adaptive mesh refinement. Summarizing, it uses fourth-order-accurate operators for the spatial derivatives in the SGS terms and in the Einstein equations (the latter are supplemented with sixth-order Kreiss-Oliger dissipation); a high-resolution shock-capturing (HRSC) method for the fluid, based on the  Lax-Friedrich flux splitting formula \cite{shu98} and the fifth-order reconstruction method MP5 \cite{suresh97}; a fourth-order Runge-Kutta scheme with sufficiently small time step $\Delta t \leq 0.4~\Delta x$ (where $\Delta x$ is the grid spacing); and an efficient and accurate treatment of the refinement boundaries when sub-cycling in time~\cite{McCorquodale:2011,Mongwane:2015}. A complete assessment of the implemented numerical methods can be found in \cite{palenzuela18,vigano19}.

The binary is evolved in a cubic domain of size $\left[-1228,1228\right]$~km. The inspiral is fully covered by seven Fixed Mesh Refinement (FMR) levels. Each consists of a cube with twice the resolution of the next larger one. In addition, we use one Adaptive Mesh Refinement (AMR) level, covering the regions where the density is above $5 \times 10^{12}~\rm{g~cm^{-3}}$ and providing a uniform resolution throughout the shear layer. With this grid structure, we achieve a maximum resolution of $\Delta x_{min} = 60$ m covering at least the most dense region of the remnant.  

\subsection{EoS and conversion to primitive variables} \label{sec:EoS}

We consider a hybrid EoS during the evolution, with two contributions to the pressure. For the cold part, we use a tabulated version of the piecewise polytrope fit to the APR4 zero-temperature EoS~\cite{read09}, with a modification to prevent superluminal speeds~\cite{Endrizzi2016}. Thermal effects are modeled for simplicity by an additional ideal gas contribution $p_{th} = (\Gamma_{th} - 1) \rho \epsilon$, with adiabatic index $\Gamma_{\rm th}= 1.8$. Within this EoS, the remnant does not collapse to a black hole within the timespan of the simulation, up to $110$ ms after the merger.

The conversion from the evolved or conserved fields to the primitive or physical ones is performed by using the robust procedure introduced in~\cite{kastaun20}, which gave us excellent results in our previous works~\cite{palenzuela22,aguilera22}. Following a common practice in GRMHD simulations, the surrounding regions of the neutron stars are filled with a relatively tenuous, low-density {\it atmosphere}, which is necessary to prevent the failure of the HRSC schemes usually employed to solve the MHD equations. To minimize unphysical states of the conserved variables outside the dense regions, produced by the numerical discretization errors of the evolved conserved fields, we set such atmosphere with a low value of $6 \times 10^{4}~\rm{g~cm^{-3}}$ (i.e., one order of magnitude lower than in~\cite{palenzuela22}, and more than ten orders of magnitude lower than the maximum values of density).

In addition, we apply the SGS terms in the regions where the density is higher than $2 \times 10^{11}~\rm{g~cm^{-3}}$ (i.e., two orders of magnitude lower than in~\cite{palenzuela22}) in order to avoid possible spurious effects near the stellar surface and in the atmosphere. Since the remnant's maximum density is above $10^{15} ~\rm{g~cm^{-3}}$, the SGS model is applied to a considerable volume including the most relevant regions of the stars and the remnant.

\subsection{Initial conditions}

The initial data is created with the {\sc Lorene} package~\cite{lorene}, using the APR4 zero-temperature EoS described above. We consider an equal-mass BNS in quasi-circular orbit, with an irrotational configuration, a separation of $45$ km and an angular frequency of $1775\ \rm{rad~s^{-1}}$. The chirp mass $M_{chirp}=1.186~M_{\odot}$ is the one inferred in the refined analysis of GW170817 \cite{LVC-170817properties}, implying a total mass $M=2.724~M_{\odot}$ for the equal mass case. 

Each star initially has a purely poloidal dipolar magnetic field that is confined to its interior, calculated from a vector potential $A_ {\phi} \propto R^2 {\rm max}(P - P_{cut},0)$, where $P_{cut}$ is a hundred times the pressure of the atmosphere, and $R$ is the distance to the axis perpendicular to the orbital plane passing through the center of each star. The maximum magnetic field (at the centers) is $5 \times 10^{11}$ G, 
several orders of magnitude lower than the large initial fields used in other simulations (e.g.,~\cite{kiuchi15,ruiz16,kiuchi18,ciolfi2019,ciolfi2020collimated,ruiz2020}) and not too far from the upper range of the oldest known NSs (millisecond pulsars and low-mass X-ray binaries~\cite{bahramian23}). Nevertheless, the initial topology is quickly forgotten after merging: acting only as a seed for the KHI, it has a negligible effect on the final magnetic field configuration of the remnant, as long as the initial values are not too large ($B\lesssim 10^{13}$ G) and the scheme and resolution are able to capture the turbulent amplification mechanisms, as it was shown in~\cite{aguilera22}.

\subsection{Analysis quantities}

We use several integral quantities to monitor the dynamics in different regions: averages of the magnetic field strength, the fluid angular velocity $\Omega \equiv \frac{d\phi}{dt} = \frac{u^{\phi}}{u^t}$ (where $u_a \equiv W (-\alpha, v_i )$ denotes the fluid four-velocity), and the plasma beta parameter, $\beta = \frac{2 P}{B^2}$. The averages for a given quantity $q$ over a certain region ${\cal N}$ will be denoted generically by:
\begin{eqnarray}
{<}q{>}_a^b = \frac{\int_{{\cal N}} q \, d{\cal N}}{\int_{{\cal N}} d{\cal N}} ,
\end{eqnarray}
where ${\cal N}$ stands for a volume $V$, a surface $S$, or a line $\ell$, and the integration is restricted to regions where the mass density is within the range $(10^a, 10^b)\, {\rm g/cm^3}$. If $b$ is omitted, it means no upper density cut is applied. From here after we will denote the \textit{bulk} as the densest region of the remnant with densities $\rho > 10^{13}~g~cm^{-3}$, and the \textit{envelope}  as the region satisfying $5 \times 10^{10} ~g~cm^{-3} < \rho < 10^{13} ~g~cm^{-3}$. In particular, we define averages over the bulk of the remnant as ${<}q{>}_{13}$, and, abusing a bit of the notation for simplicity, averages over the envelope as ${<}q{>}^{13}_{10}$. Surface integrals are carried out over a cylinder $S$ with axis passing through the center of mass and orthogonal to the orbital plane. Line integrals are carried out over a circle $\ell$ centered at the center of mass and orthogonal to the orbital plane. We also compute global quantities, integrated over the whole computational domain, such as the total magnetic energy $E_{mag}$, thermal energy $E_{th}$ and rotational kinetic energy $E_{rot}$, as defined in~\cite{palenzuela22}.

In addition, we compute the spectral distribution of the kinetic and magnetic energies over the spatial scales. For the magnetic spectra, we also calculate the poloidal and toroidal contributions separately. Further details of the numerical procedure to calculate the spectra can be found in~\cite{aguilera2020,vigano19,vigano20,palenzuela22}. With these spectrum distributions we can define the spectra-weighted average wave-number
\begin{equation}
\langle k \rangle \equiv \frac{\int_k k\,{\cal E}(k) \,dk} {\int_k {\cal E}(k)\, dk}~,
\end{equation}
with an associated length scale $\langle L \rangle = 2\pi/ \langle k \rangle $  which represents the typical \textit{coherent} scale of the structures present in the field.

Finally, we here introduce the \textit{non-axisymmetric intensity}, a new topological indicator which might be useful to estimate the amount of turbulence in some scenarios. We can decompose any given vector field $U^i$ into its average and a residual component,
\begin{equation}
	U^i = \overline{U}^i + {\delta U}^i ~~.
\end{equation}
The remnant is characterized by a growing axial symmetry, such that the flow becomes mostly dominated by the azimuthal component of the vector fields. Then, for each distance $R$ to the center of mass, we consider the averages of this azimuthal component over the circle in the orbital plane $z_{orb}$, namely
\begin{eqnarray}
	\overline{U}_{\phi}(R)&=& \frac{1}{2 \pi} \int_0^{2\pi} U_{\phi}(R,\phi,z_{orb})~d\phi ~~,~~
	\\
	\delta U_{\phi}(R) &=& \frac{1}{2 \pi} \int_0^{2\pi} \left|U_{\phi}(R,\phi,z_{orb}) - \overline{U}_{\phi}(R) \right|~d\phi ~.
\end{eqnarray}
We then define the axisymmetric energy contributions of the average and the residuals as follows:
\begin{eqnarray}
	 \overline{E}(R) = \overline{U}^\phi \overline{U}_\phi~~~,~~~
	{\delta E}(R) = {\delta U}^\phi {\delta U}_{\phi}~,
\end{eqnarray}
which allow us to calculate the following magnetic and kinetic indicators (i.e., using $U^i=B^i$ and $U^i= \sqrt{\rho} v^i$ respectively) as the fraction of non-axisymmetric contributions to the total energy, that is
\begin{equation}\label{eqn:Iturbulence}
	I_{kin}(R) = \frac{{\delta E}_{kin}}{{\bar E}_{kin} + {\delta E}_{kin}}
	~~,~~
	I_{mag}(R) = \frac{{\delta E}_{mag}}{{\bar E}_{mag} + {\delta E}_{mag}} ~.
\end{equation}
Although we are taking into account only the toroidal contribution to the energy, which is the dominant one, we have checked that similar values are obtained when the other components are also included. These axisymmetric indicators are zero for perfect axial symmetry and approach one if the residuals are dominating. Note that with these definitions the residuals can either have a turbulent origin (small scales), or can consist of large/intermediate non-axisymmetric modes. Therefore, we infer how relevant is the turbulence by looking simultaneously at the axisymmetric indicators, the spectra and, visually, the orbital plane slices.

\section{Results} \label{sec:results}

Several MHD processes operate during and after the merger of the stars to amplify and sustain the magnetic field in timescales up to hundreds of milliseconds. The most relevant ones that have been proposed for this scenario include the KHI, the MRI and the winding mechanism. First, the shear layer that forms when the surfaces of the NSs touch is subject to the KHI, which generate a turbulent state that can lead to a small-scale dynamo. Secondly, the strong differential rotation tends to wind up the magnetic field lines, shearing the poloidal component to amplify the toroidal one. The MRI~\cite{balbus91,balbus98} could also amplify the magnetic field if an ordered, large-scale field is present (late stages of post-merger), 
though a firm evidence in BNS mergers with realistic, turbulent magnetic topology is still lacking (see \cite{palenzuela22} for a detailed interpretation). These processes will be the subject of our discussion in the next section, after presenting the results of our simulations.

\subsection{Qualitative dynamics}

\begin{figure*}
	\centering
	\includegraphics[width=0.273\linewidth, trim={6.5cm 4.5cm 0 0}, clip]{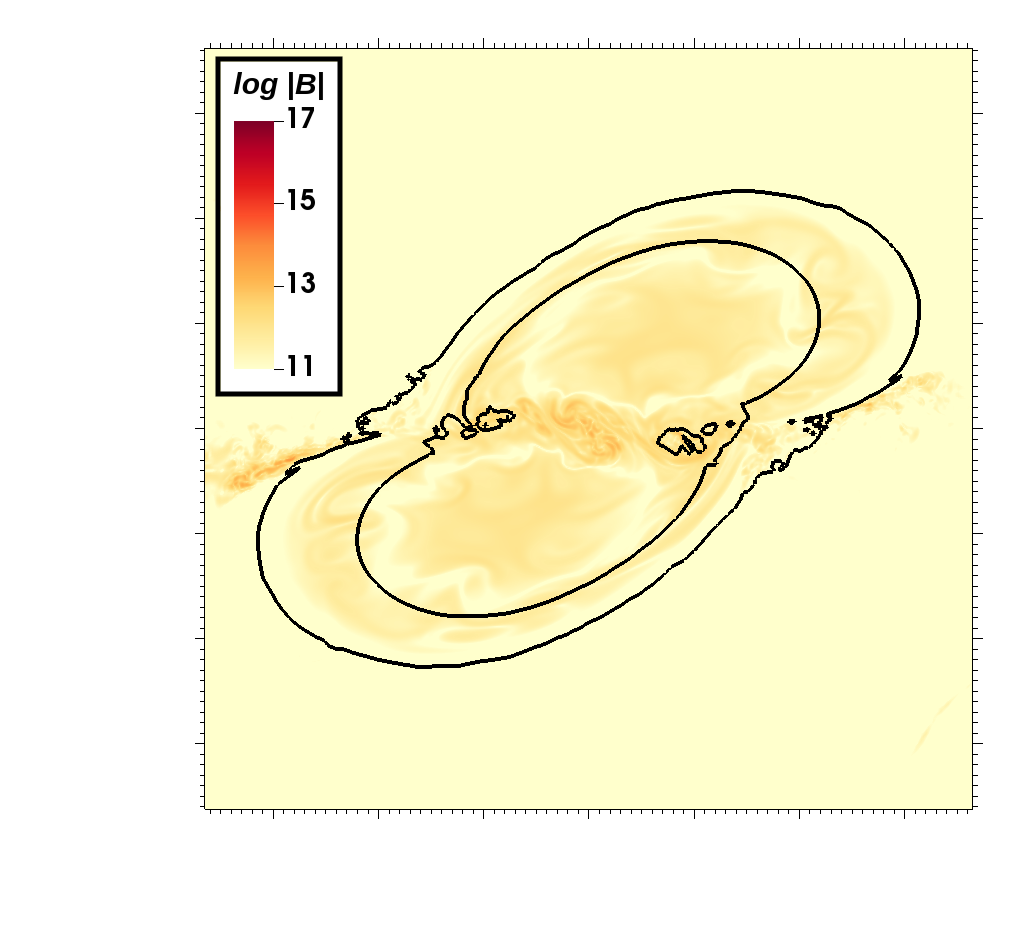}
	\includegraphics[width=0.273\linewidth, trim={6.5cm 4.5cm 0 0}, clip]{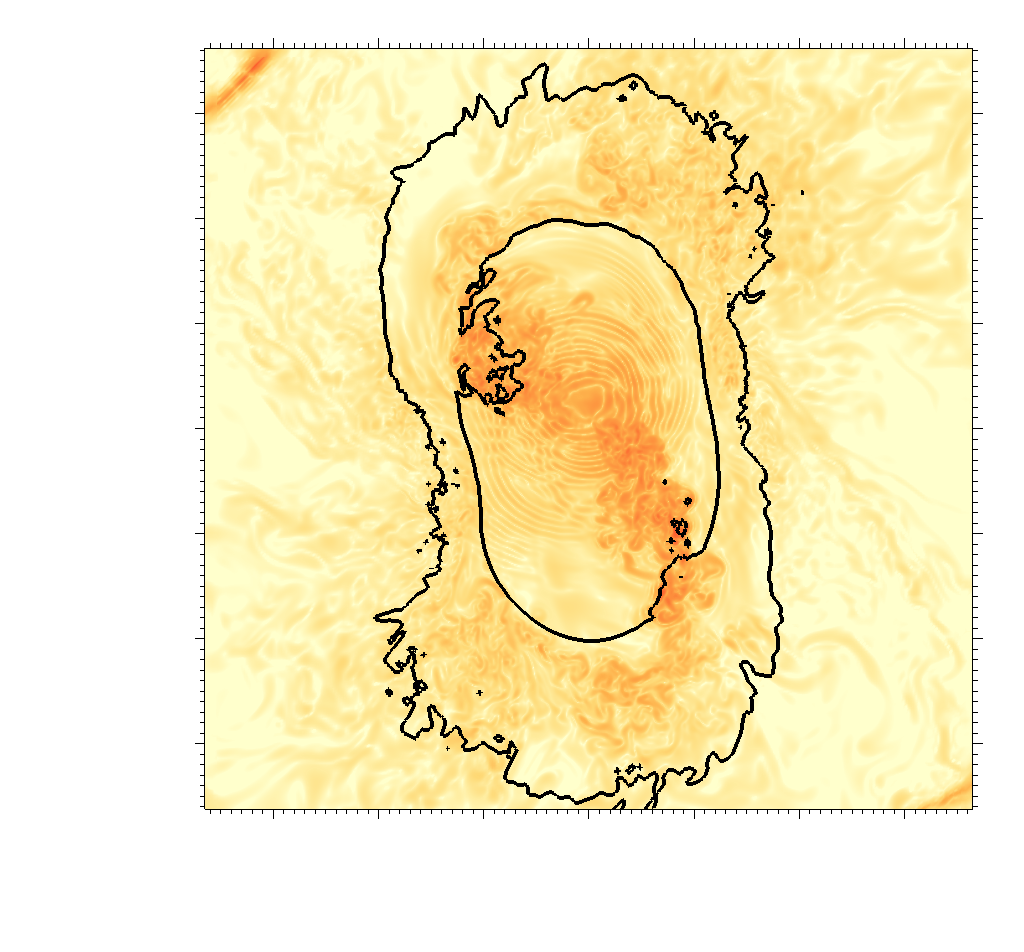}
	\includegraphics[width=0.273\linewidth, trim={6.5cm 4.5cm 0 0}, clip]{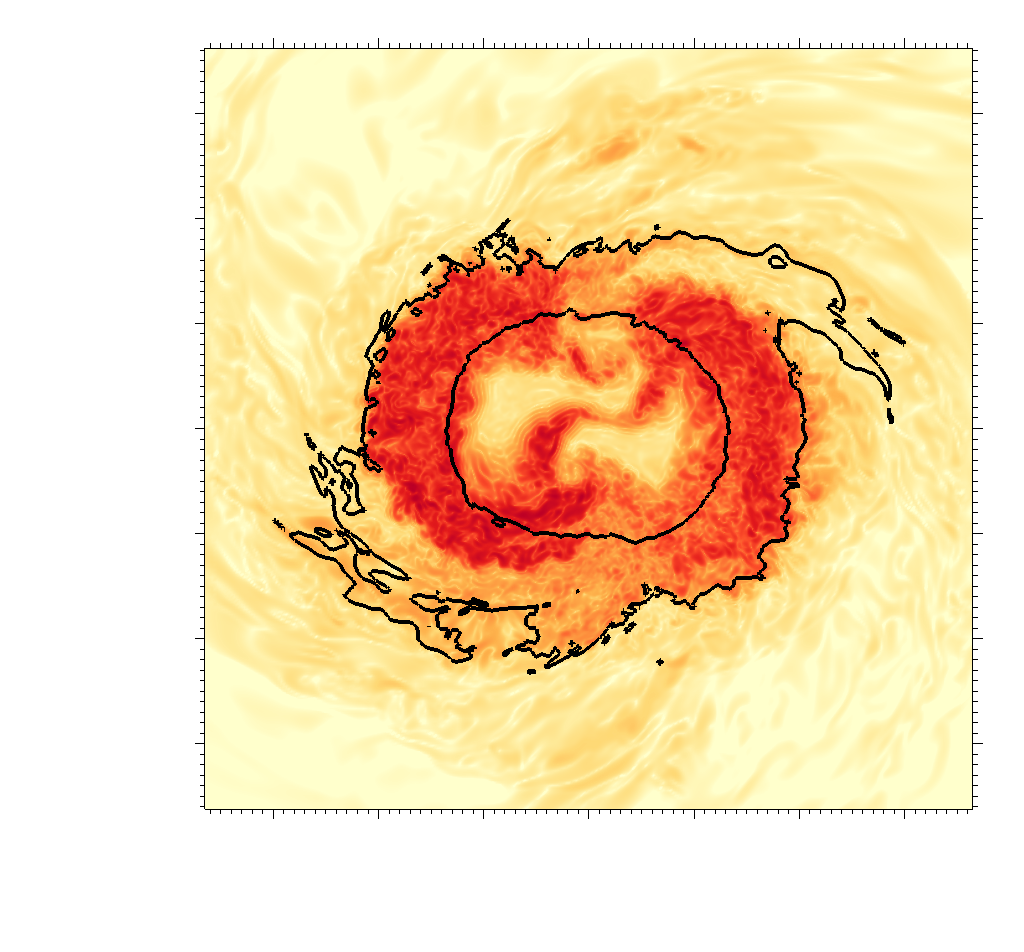}\\
	\includegraphics[width=0.273\linewidth, trim={6.5cm 4.5cm 0 0}, clip]{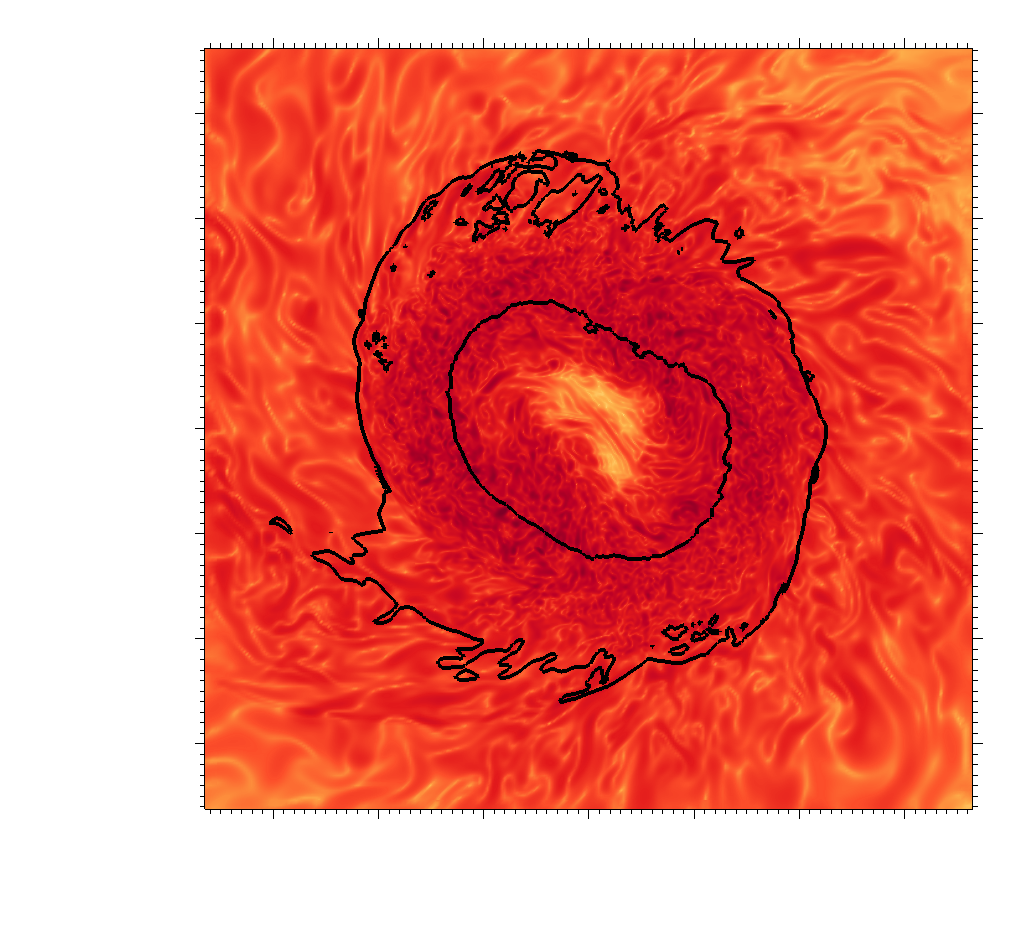}
	\includegraphics[width=0.273\linewidth, trim={6.5cm 4.5cm 0 0}, clip]{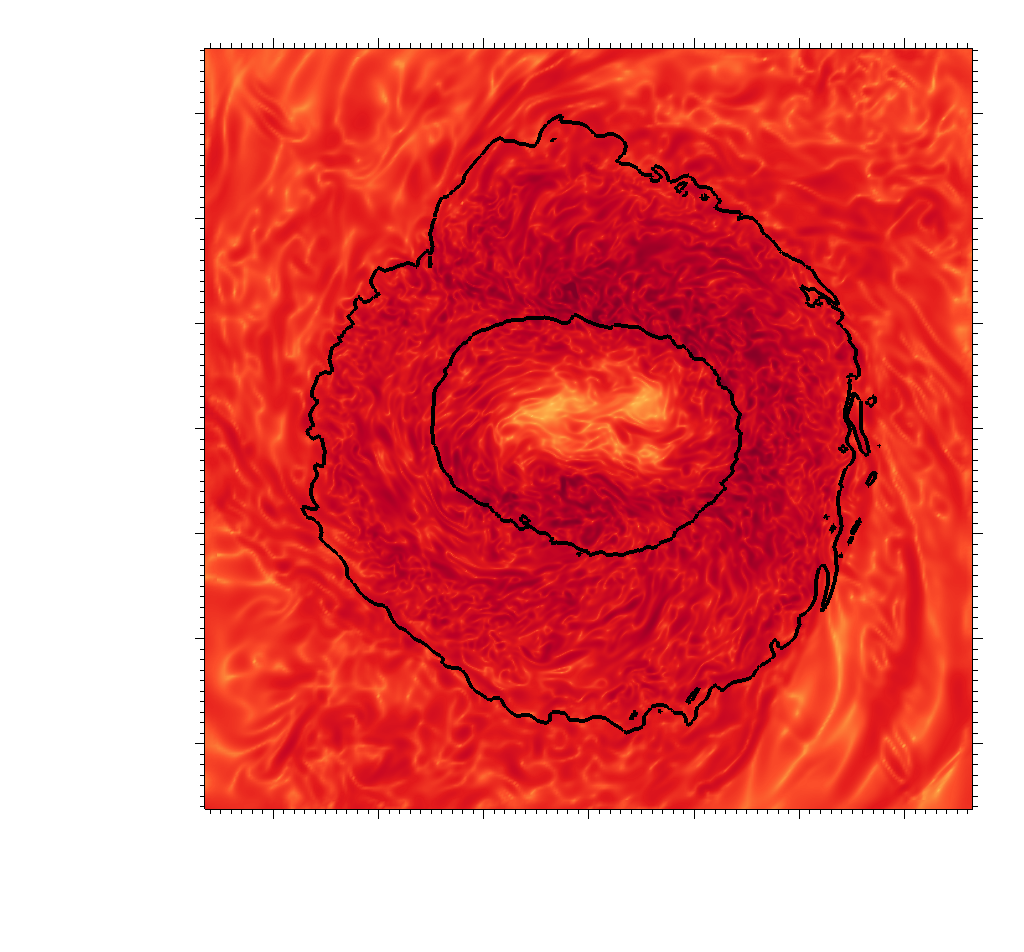}
	\includegraphics[width=0.273\linewidth, trim={6.5cm 4.5cm 0 0}, clip]{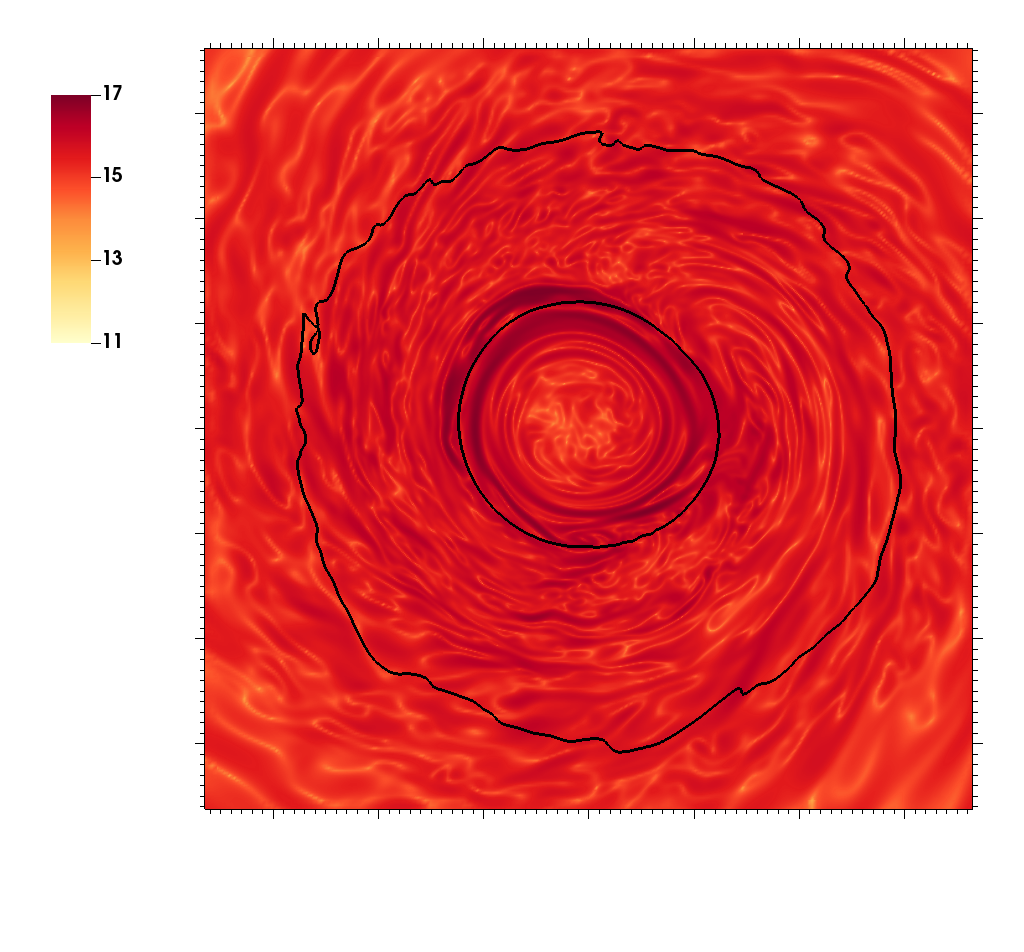}\\
	\includegraphics[width=0.273\linewidth, trim={6.5cm 2cm 0 0}, clip]{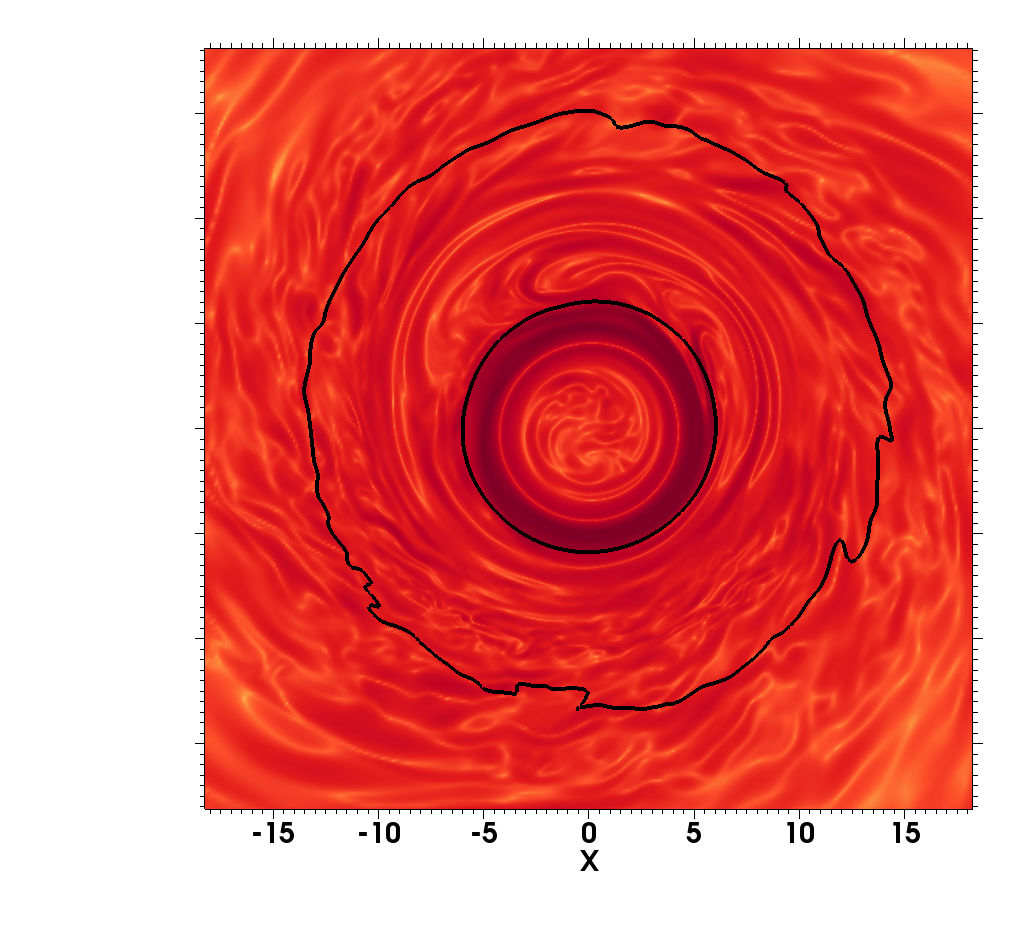}
	\includegraphics[width=0.273\linewidth, trim={6.5cm 2cm 0 0}, clip]{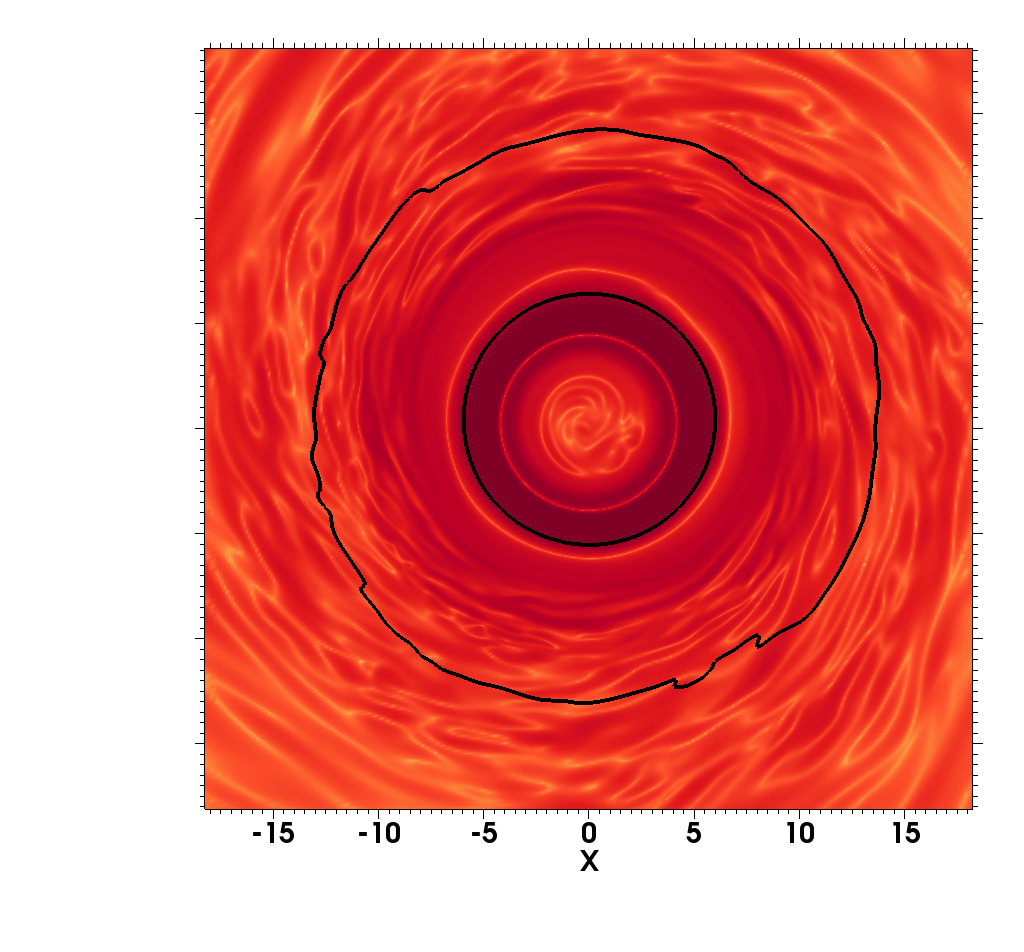}
	\includegraphics[width=0.273\linewidth, trim={6.5cm 2cm 0 0}, clip]{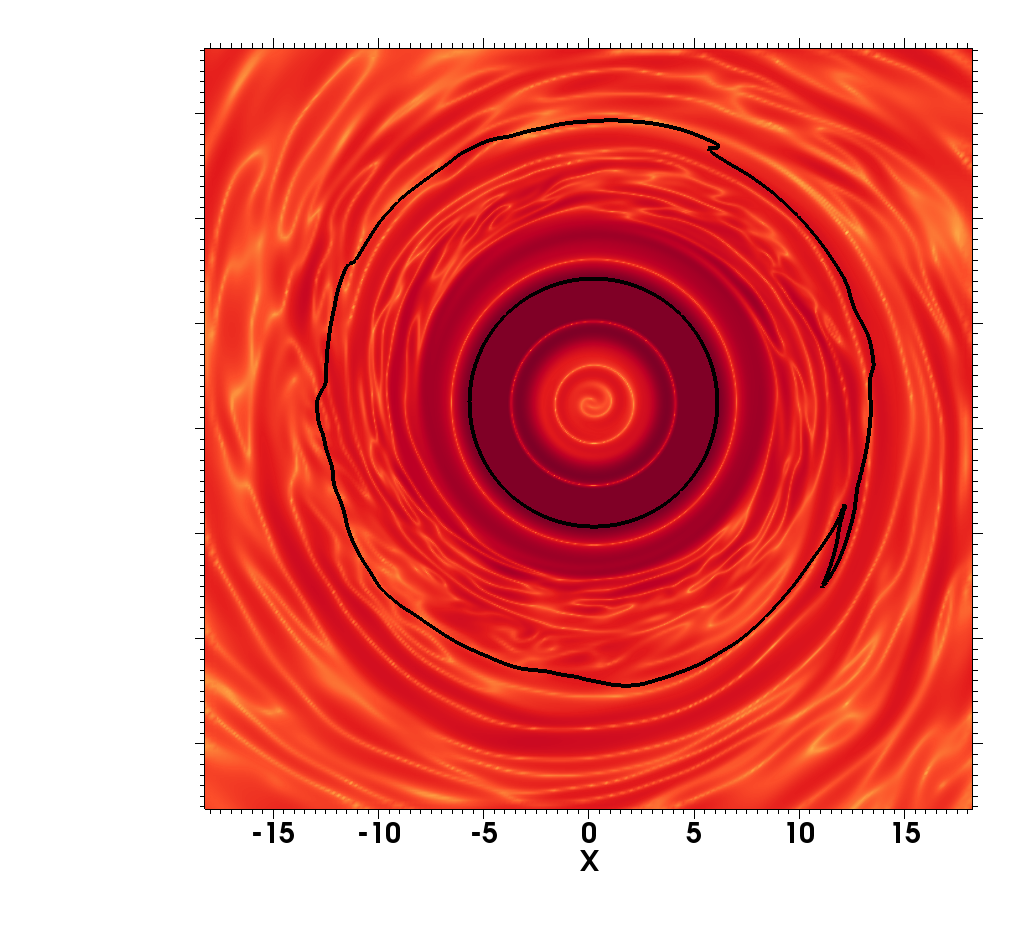}
	\caption{\textit{Magnetic field evolution in the orbital plane}. The magnetic field strength $|\vec{B}|$ is shown in logarithmic scale at times $t=\{0.5, 1.25, 2.5, 3.75, 5, 15, 25, 50, 110\}$ ms after the merger. Outer and inner black lines depict density contours of $\rho = 10^{13}~\rm{g~cm^{-3}}$ and $5 \times 10^{14}~\rm{g~cm^{-3}}$, respectively. The length is given in units of km.}
	\label{fig:slices_B2}
\end{figure*}

The BNS system merges approximately after $5$ orbits, producing a differentially rotating remnant that relaxes to an hypermassive-neutron star (HMNS) in a few milliseconds. Just before the merger occurs (hereafter, $t = 0$), we set a purely poloidal magnetic field distributed over the interior of each star with a maximum magnetic field of $5 \times 10^{11}~\rm{G}$ at their centers.

\begin{figure*}
	\centering
	\includegraphics[width=0.3\linewidth, trim={2cm 4.5cm 0 0}, clip]{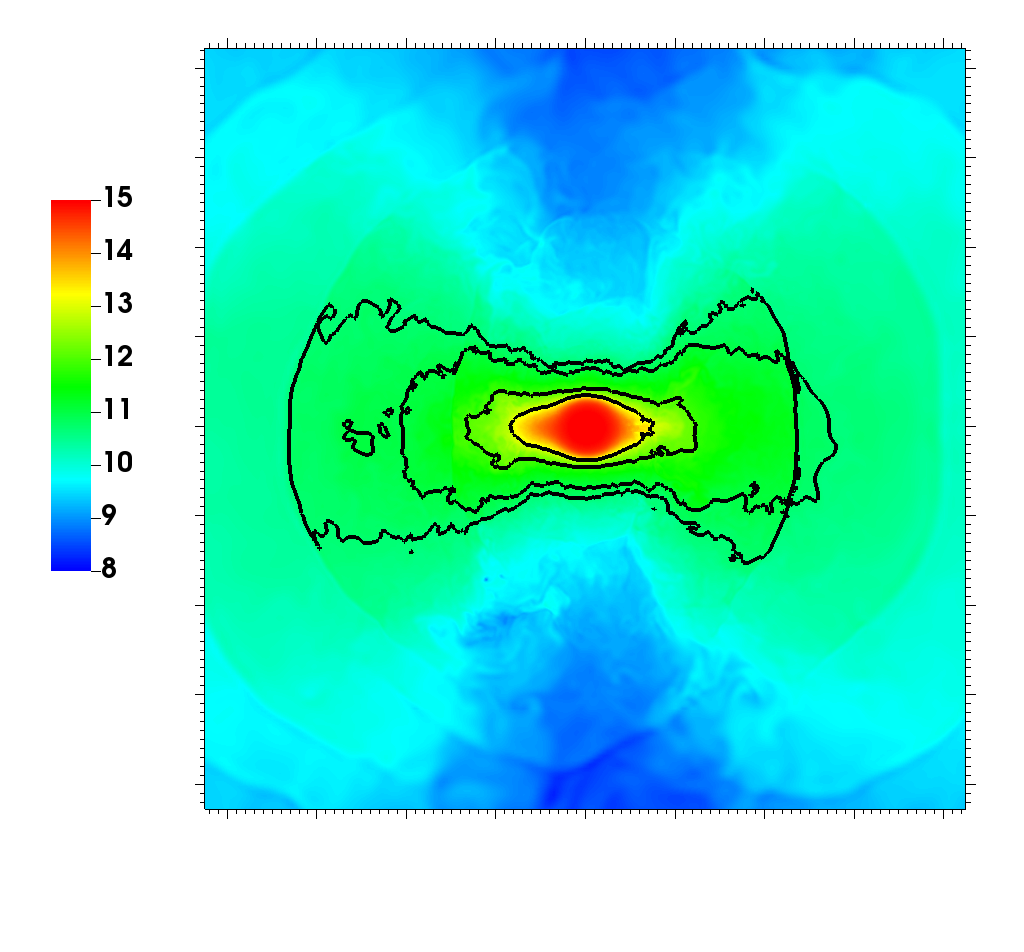}
	\includegraphics[width=0.26\linewidth, trim={6.5cm 4.5cm 0 0}, clip]{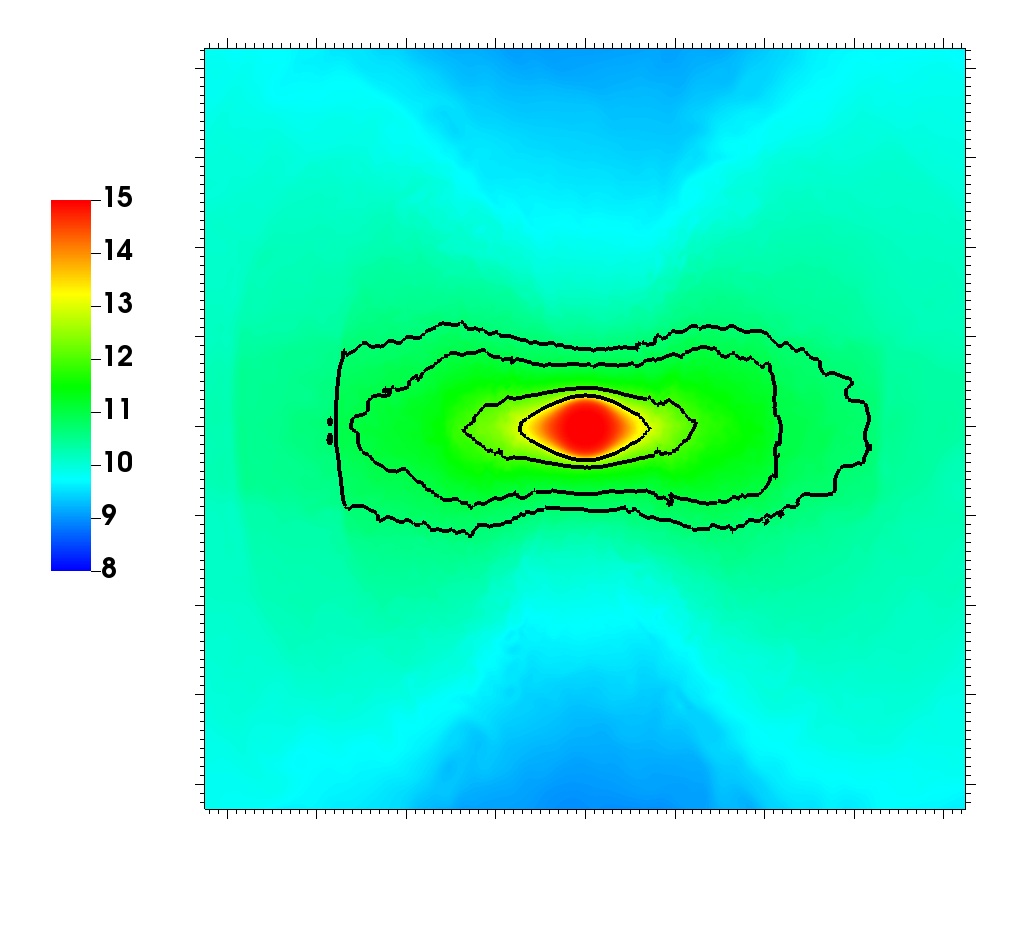}
	\includegraphics[width=0.26\linewidth, trim={6.5cm 4.5cm 0 0}, clip]{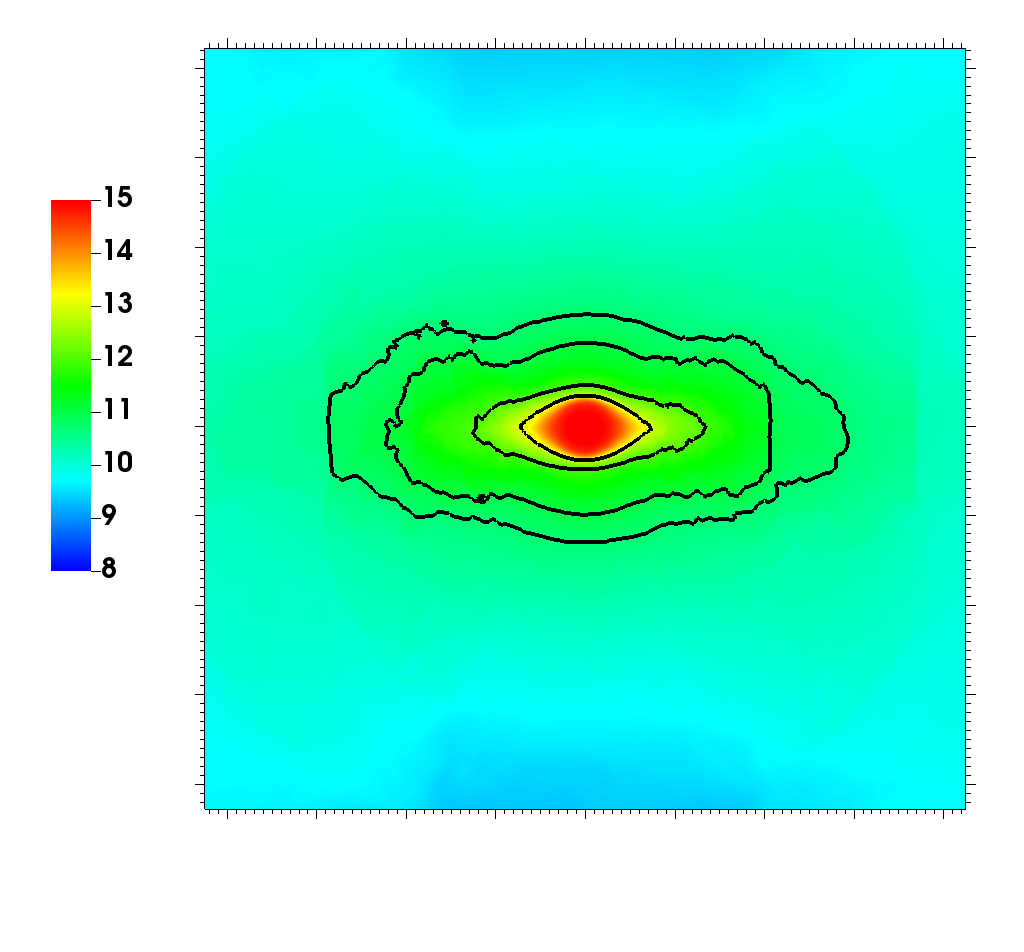}\\
	\includegraphics[width=0.3\linewidth, trim={2cm 4.65cm 0 0}, clip]{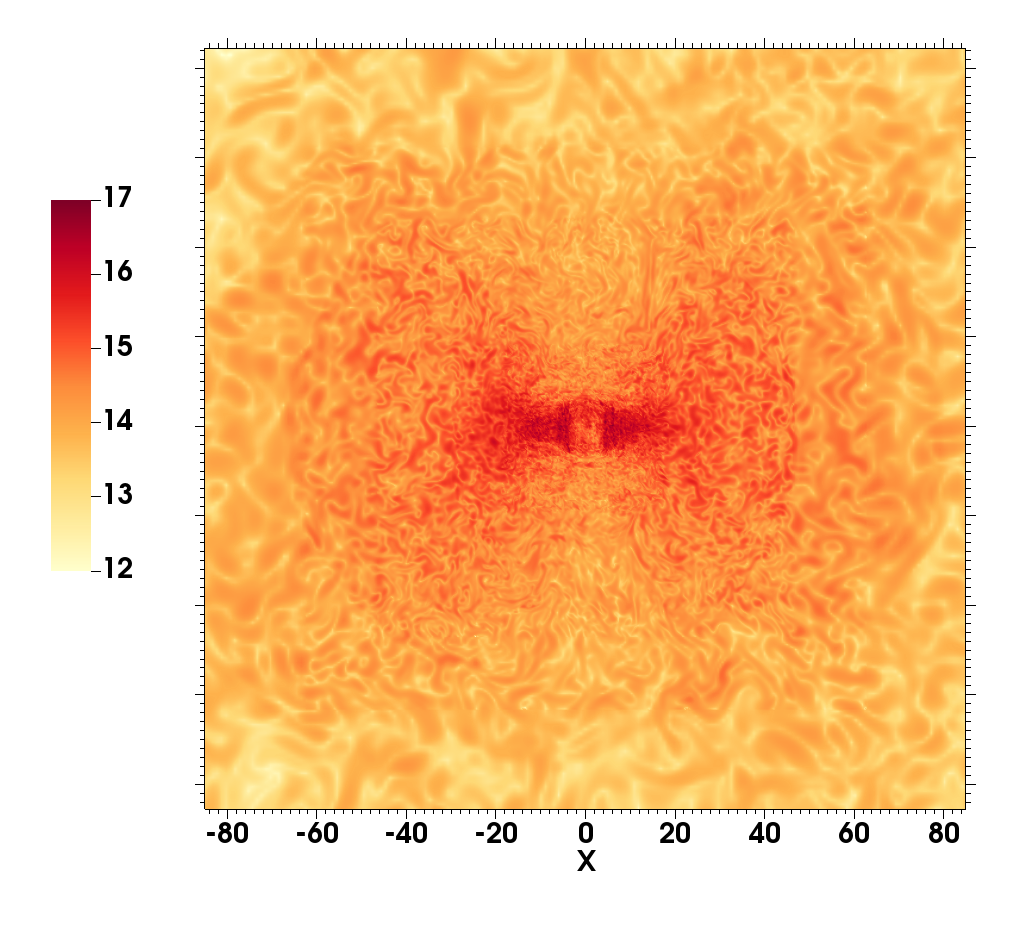}
	\includegraphics[width=0.26\linewidth, trim={6.5cm 4.65cm 0 0}, clip]{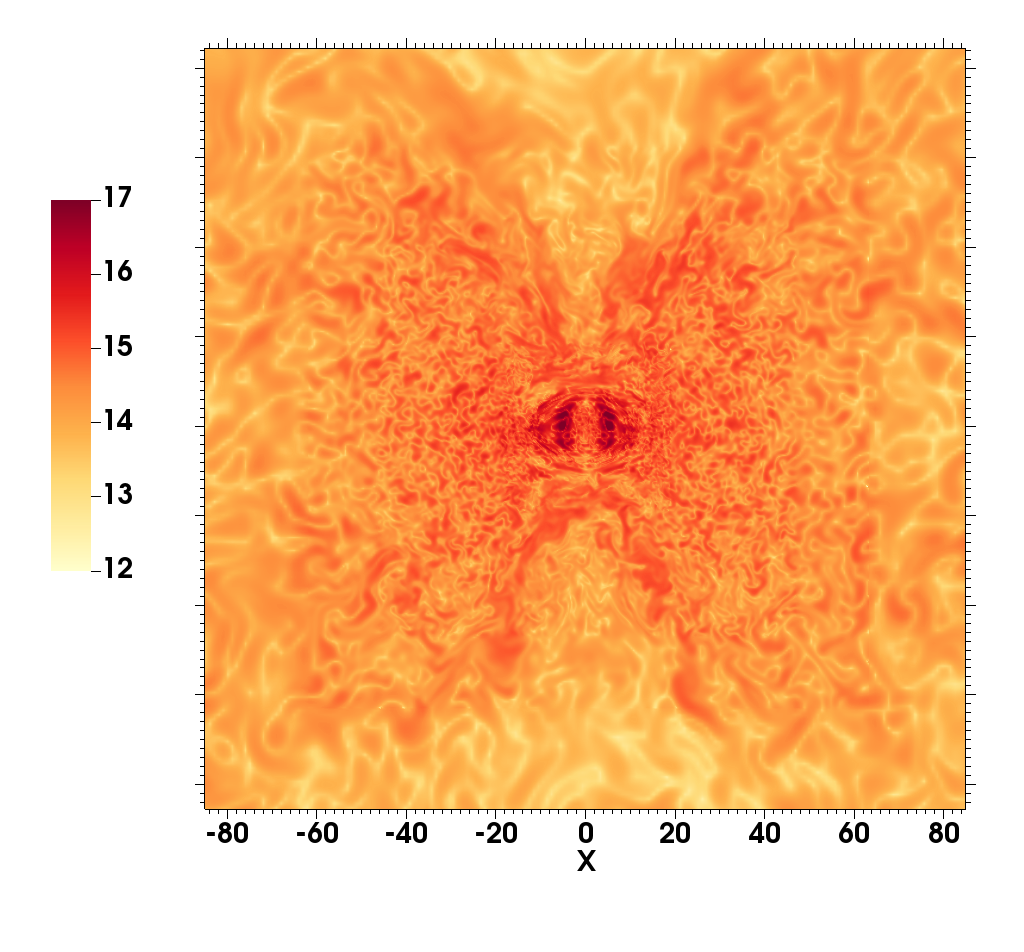}
	\includegraphics[width=0.26\linewidth, trim={6.5cm 4.65cm 0 0}, clip]{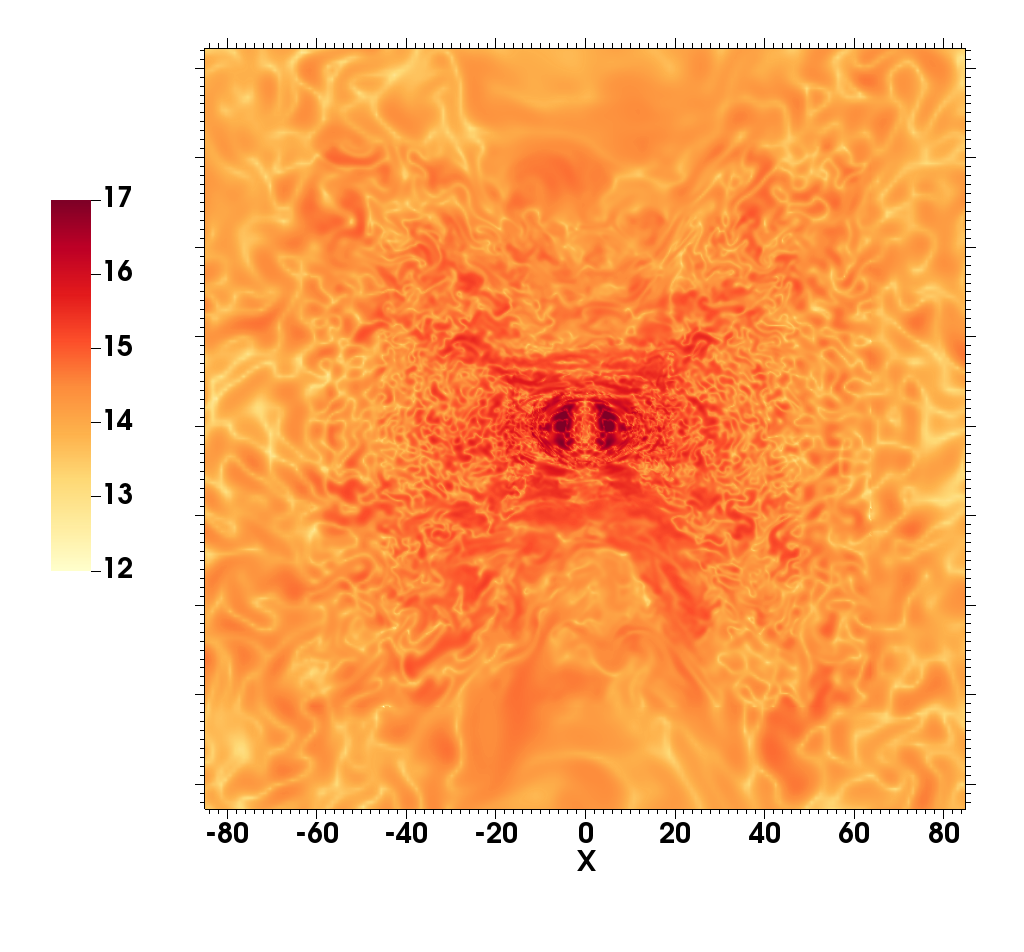}\\
	\includegraphics[width=0.3\linewidth, trim={2cm 4.5cm 0 0}, clip]{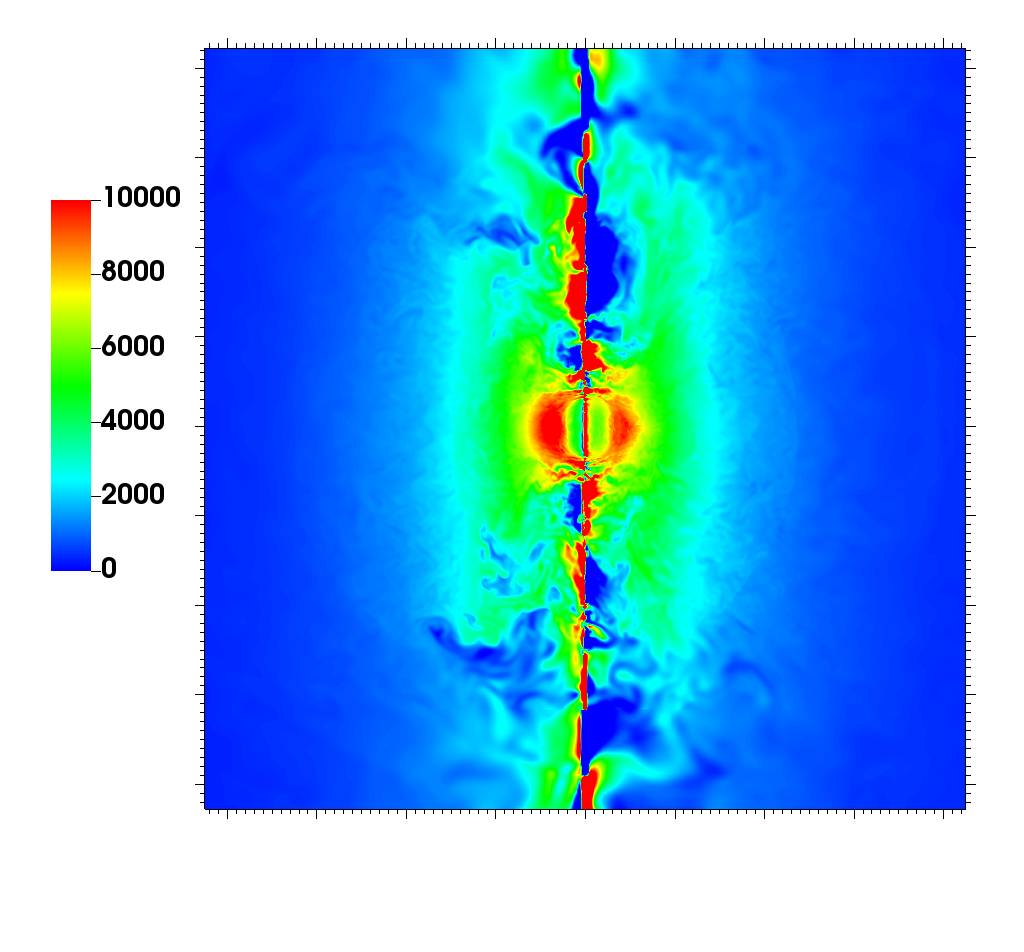}
	\includegraphics[width=0.26\linewidth, trim={6.5cm 4.5cm 0 0}, clip]{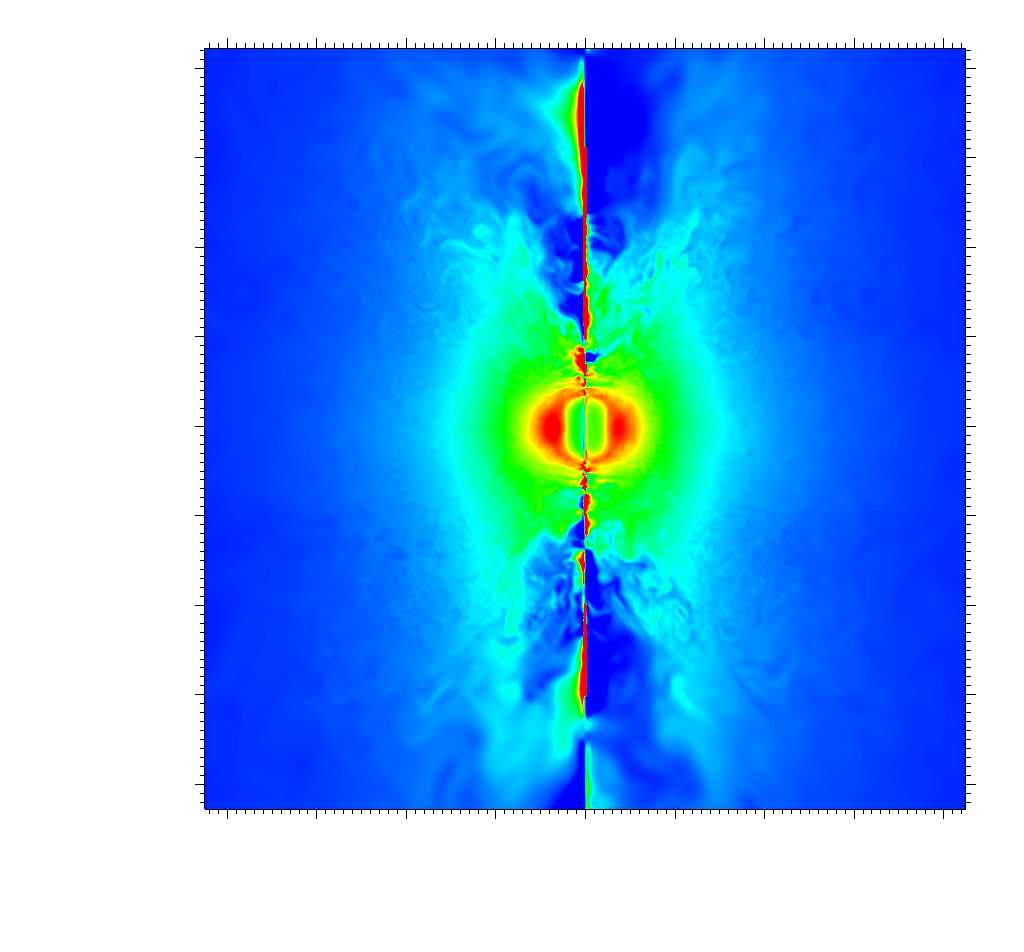}
	\includegraphics[width=0.26\linewidth, trim={6.5cm 4.5cm 0 0}, clip]{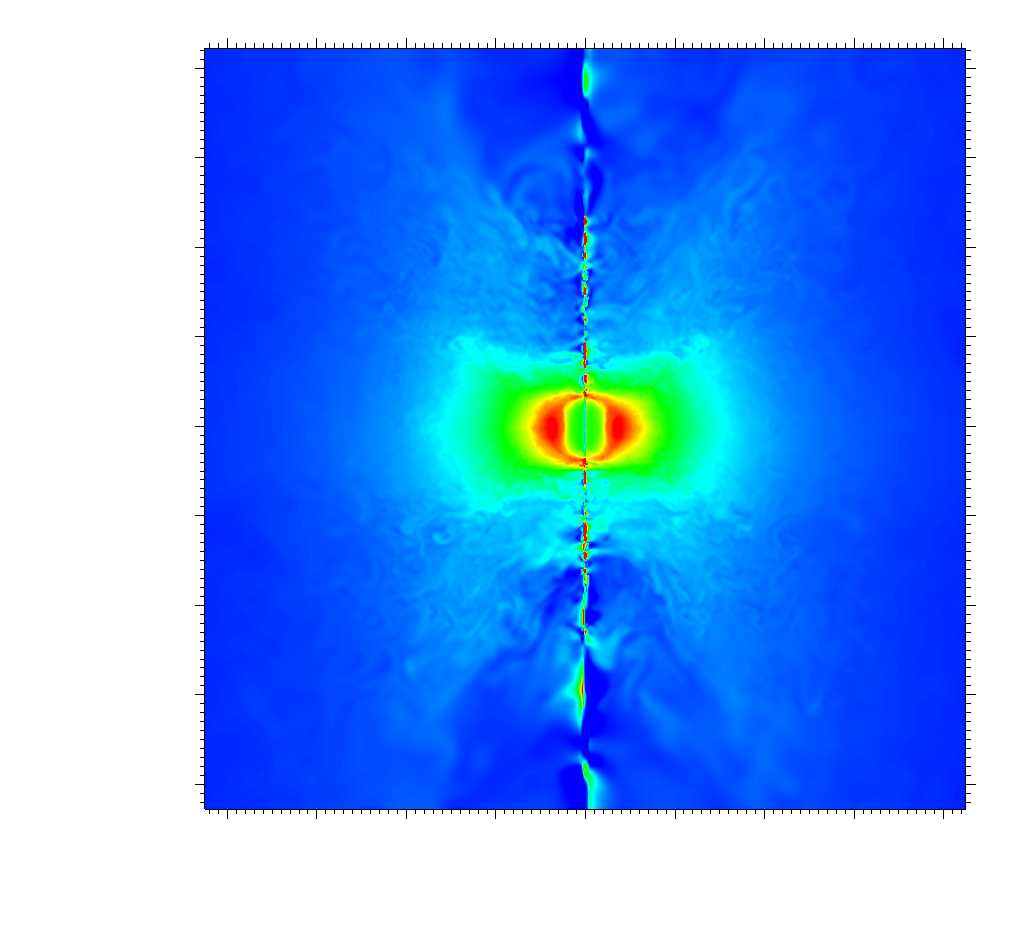}\\
	\includegraphics[width=0.3\linewidth, trim={2cm 2cm 0 0}, clip]{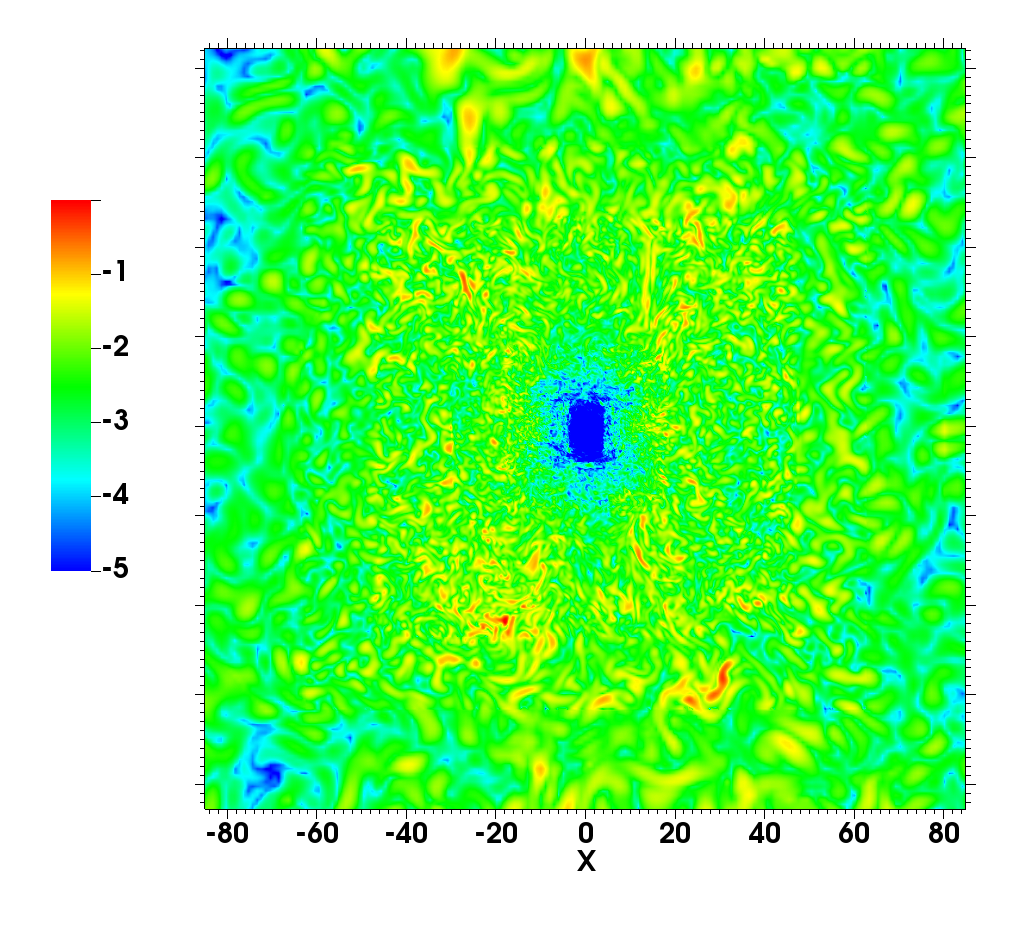}
	\includegraphics[width=0.26\linewidth, trim={6.5cm 2cm 0 0}, clip]{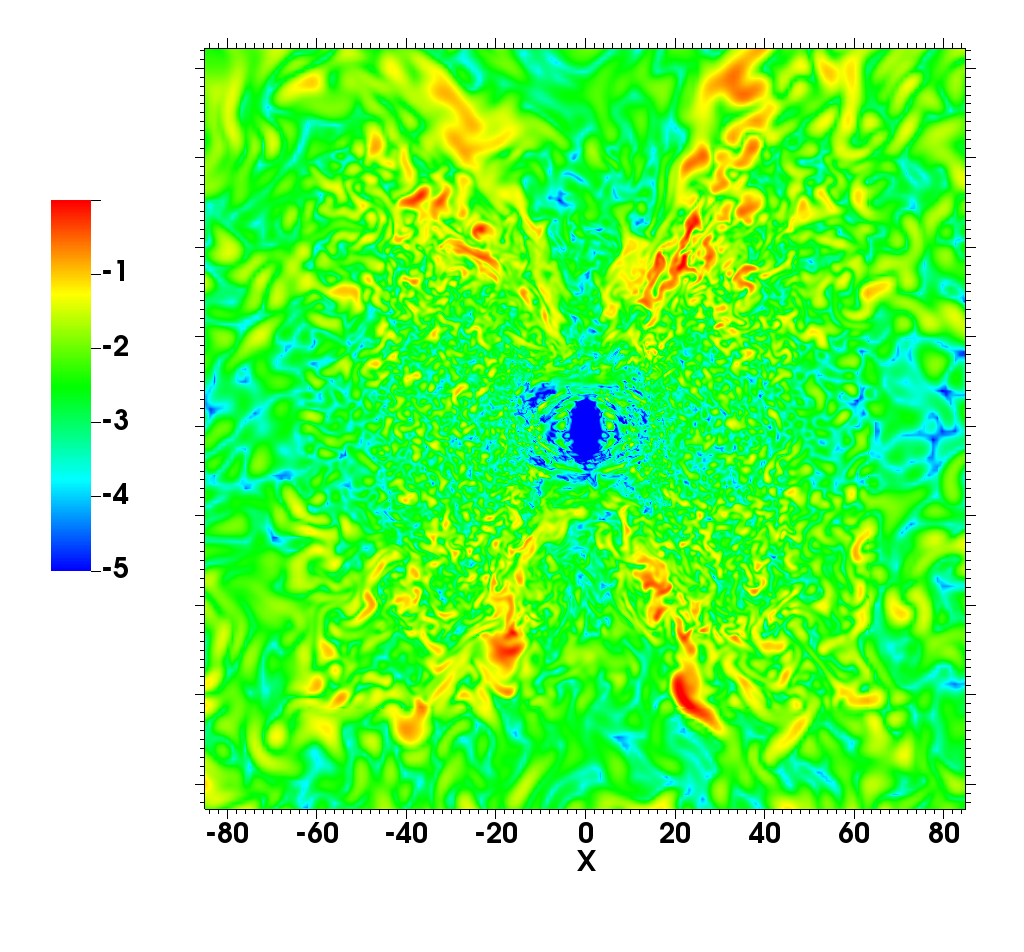}
	\includegraphics[width=0.26\linewidth, trim={6.5cm 2cm 0 0}, clip]{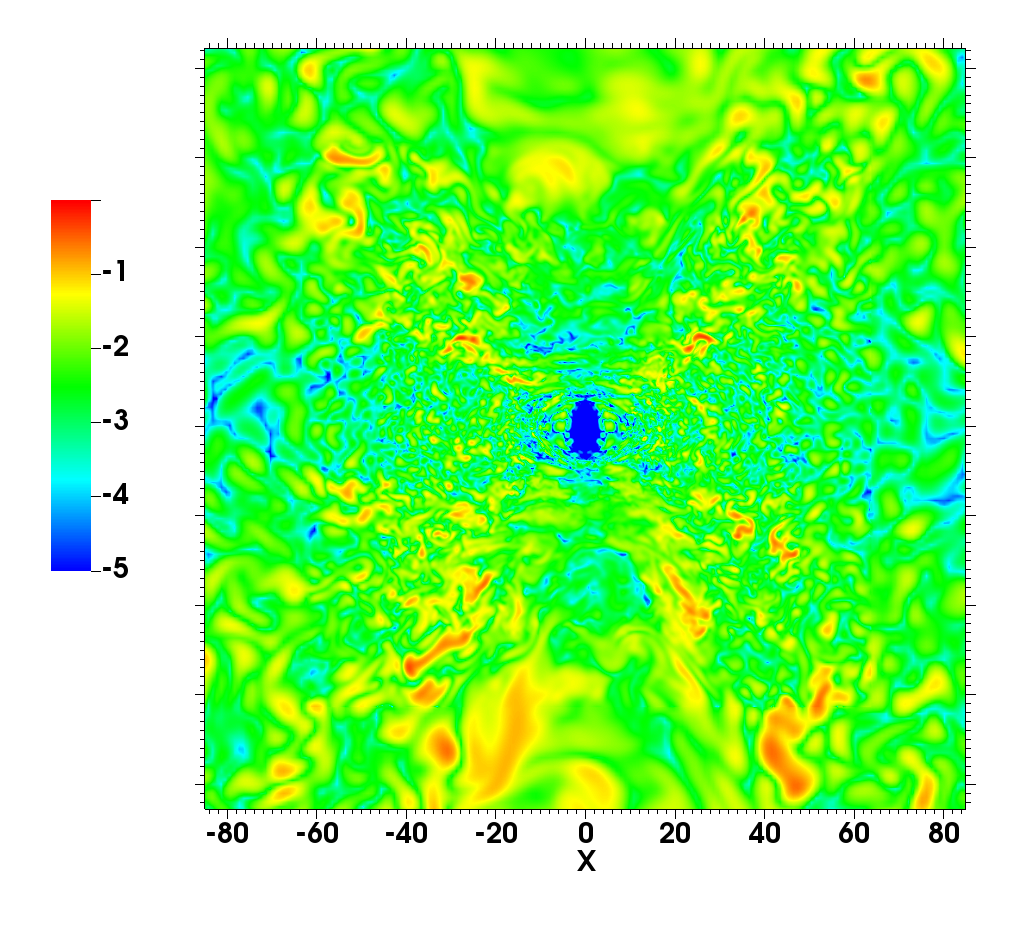}
	\caption{\textit{2D plots in the meridional plane}. The rows show, from top to bottom, the rest-mass density in [g~cm$^{-3}$], magnetic field intensity in [G], angular velocity of the fluid $\Omega$ in [rad~s$^{-1}$], and inverse of $\beta$ factor, at $t = \{10, 50, 100\}$ ms after the merger (first, second and third column, respectively). The black solid lines in the rest-mass density slices represent constant density surfaces at $\rho = \{5\times10^{10}, 10^{11}, 10^{12}, 10^{13}\}$ g~cm$^{-3}$. The length is given in km.}
	\label{fig:meridional_full}
\end{figure*}

Some representative time snapshots of the simulation, showing the merger and post-merger magnetic field dynamics in the orbital plane $z = 0$, are displayed in Fig.~\ref{fig:slices_B2}. They show how the shape of the remnant quickly restructures itself from two cores that rotate and bounce towards each other into a differentially rotating single remnant. 
The Kelvin-Helmholtz instability is produced in the shear layer of thickness $d$ between two fluids moving with opposite directions. All the modes with wavelength $\lambda > d$ are going to be unstable and produce vortexes, twisting the magnetic fields and amplifying their strength exponentially in a timescale much smaller than the characteristic timescale of the system. The KHI can be observed in the first row of the figure, with a magnetic field reaching maximum local values of $10^{17} ~\rm{G}$.
Smaller wavelengths grow faster and interact with the other eddies, inducing a turbulent state that further amplifies the magnetic energy via small-scale dynamo processes and propagates to all the remnant due to the rotation and the bouncing of the cores. 	
In the third row, at later times after the merger, the magnetic field changes from fully turbulent to a very structured shape mainly due to the winding mechanism. This seems to be one of the main processes dominating the evolution of magnetic fields for $t \gtrsim 25~\rm{ms}$, when the remnant has an almost axisymmetric structure. A very clear ring structure in the intensity develops near $\sim 5 \rm{km}$ at these late times, which will be discussed later in more detail.

\begin{figure*}[!ht]
	\begin{center}
		\includegraphics[scale=0.16]{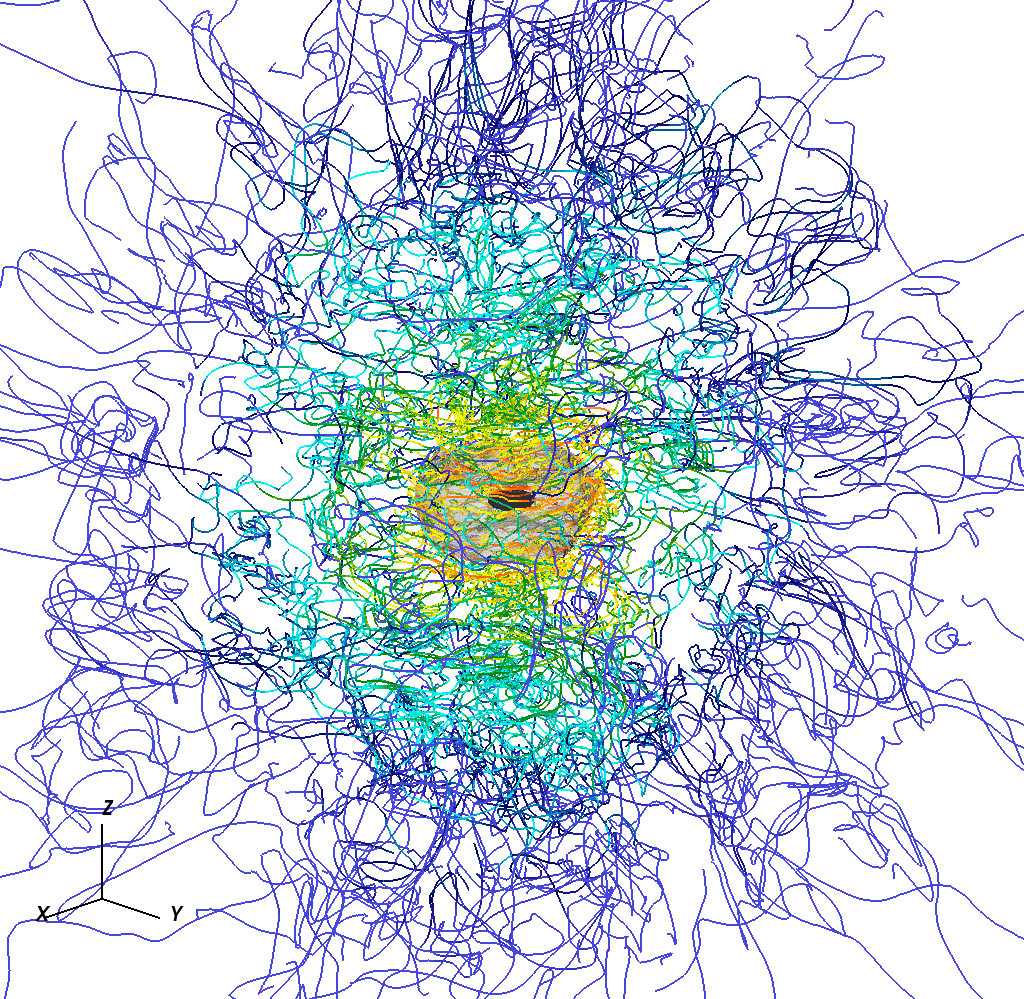}
		\includegraphics[scale=0.16]{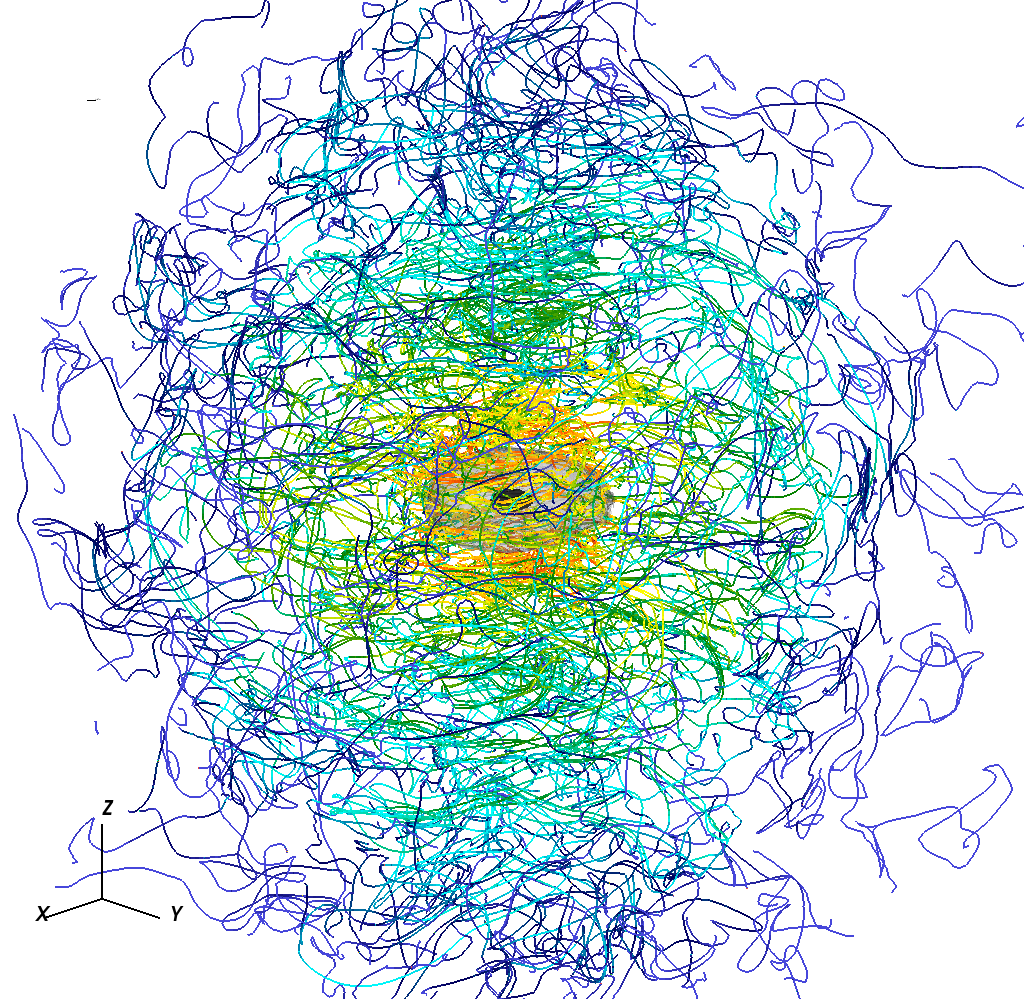}
		\includegraphics[scale=0.16]{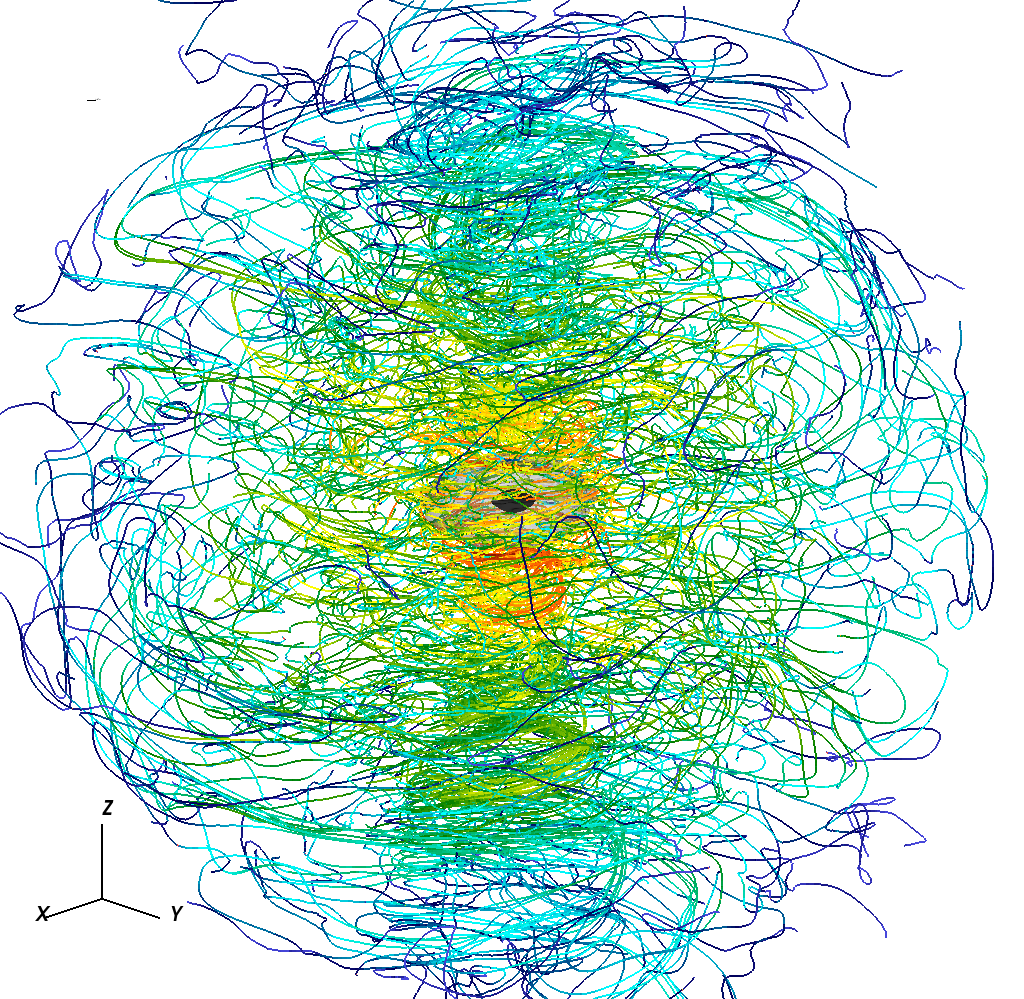}\\
		\vspace{0.5cm}
		\includegraphics[scale=0.16]{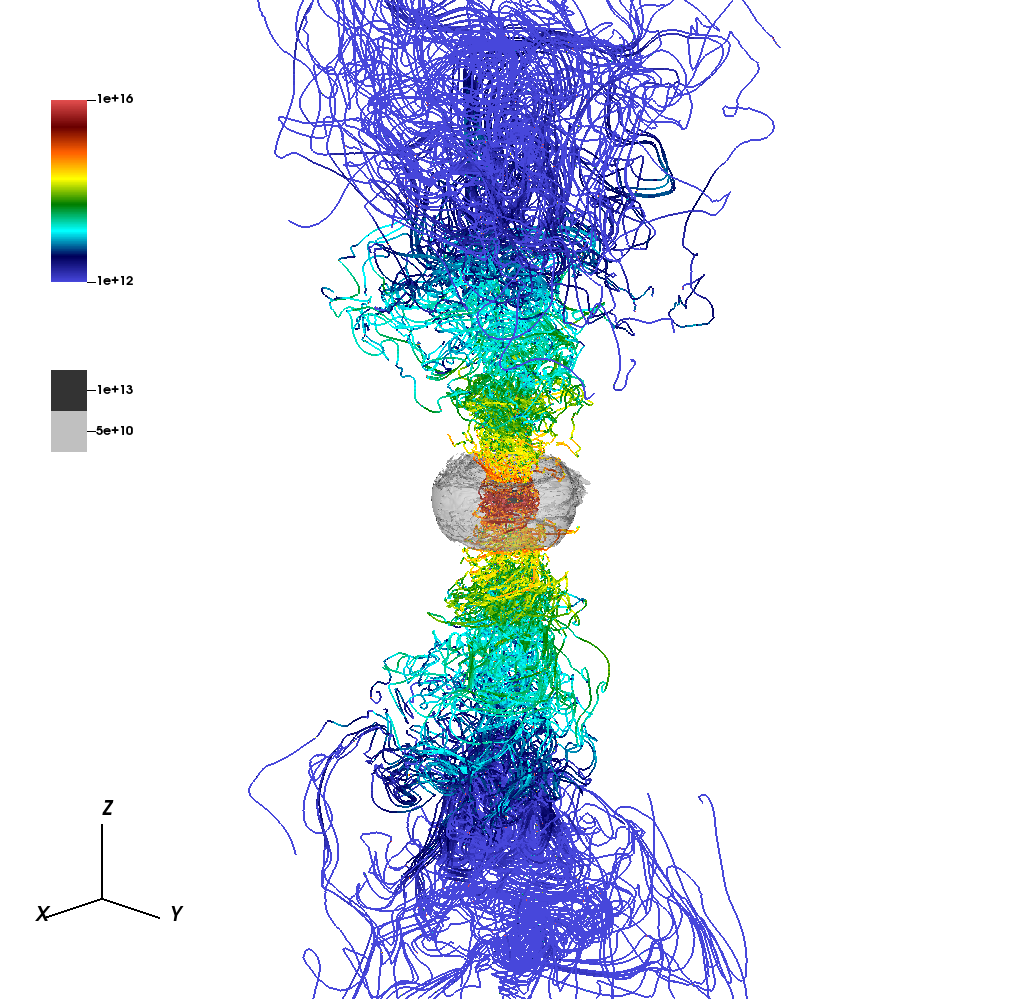}
		\includegraphics[scale=0.16]{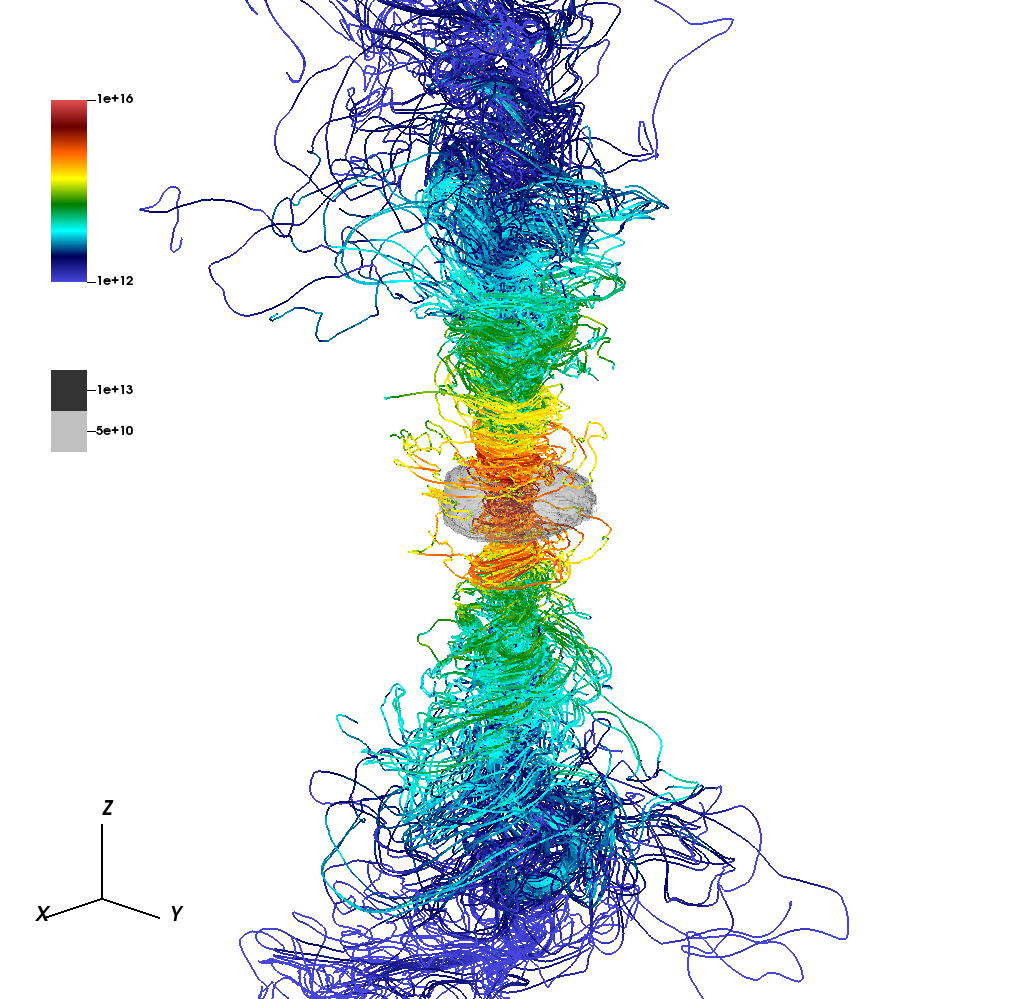}
		\includegraphics[scale=0.16]{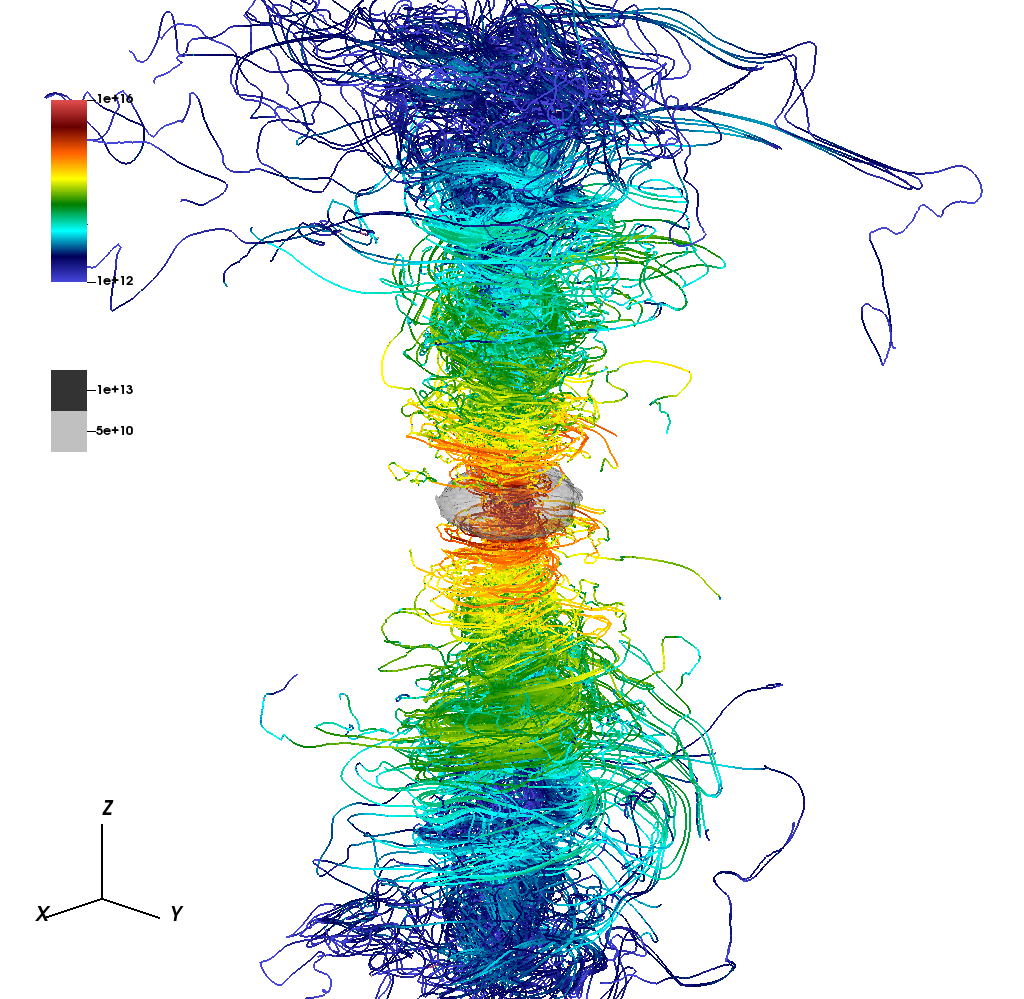}		
	\end{center}
	\caption{ \textit{Jet formation}. Streamlines of the magnetic field, at times $t = \{15, 50, 110\}$ ms after the merger (first, second and third column, respectively). The top panels employ seeds distributed isotropically for the streamline integration, while that the bottom panels have seeds near the $z$-axis. The density constant surface at $\rho = 5\times10^{10}$ g~cm$^{-3}$ is also shown. Notice the helicoidal jet-like structure developing during the post-merger phase.
	}
	\label{fig:3D_streamlines}
\end{figure*}

In Fig.~\ref{fig:meridional_full}, different relevant quantities are plotted at the meridional plane. From top to bottom, they represent, respectively: the density, the magnetic field, the angular velocity and $\beta^{-1}$ at $t = (10, 50, 100)~\rm{ms}$ after the merger. The density plots include solid lines that represent isodensity surfaces at $\rho = \{5 \times 10^{10}, 10^{11}, 10^{12}, 10^{13}\}$~g~cm$^{-3}$. The density and the angular velocity slices show that the remnant strongly approaches axial symmetry at late times, especially in its densest part. The angular velocity grows from its value at the inner region to a peak value at $\sim 7~\rm{km}$, and then decays quickly in the meridional direction and more softly in the radial one. Matter is filling the regions near the $z$-axis, above and below the remnant, which barely rotates at $100~\rm{ms}$ after the merger. The $\beta^{-1}$ slices show that the most of the remnant is governed by the fluid pressure. 

The streamlines of the magnetic field lines are displayed in Fig.~\ref{fig:3D_streamlines} at $t=\{15, 50, 110\}~\rm{ms}$ after the merger, where the gray/black contours represent constant density surfaces. We can see that an helicoidal structure, with a high intensity near the bulk of the remnant, is gradually being formed at times $\gtrsim 50~\rm{ms}$.  The helicoidal structure seems to be a necessary condition for a jet formation. However, this part of the remnant is still matter dominated (as seen in the $\beta^{-1}$ slices from Fig.~\ref{fig:meridional_full}), probably due to the lack of neutrinos to clear the funnel near the $z$-axis~\cite{cusinato2022,mosta2020}.

\begin{figure}
	\centering
	\includegraphics[width=0.9  \linewidth]{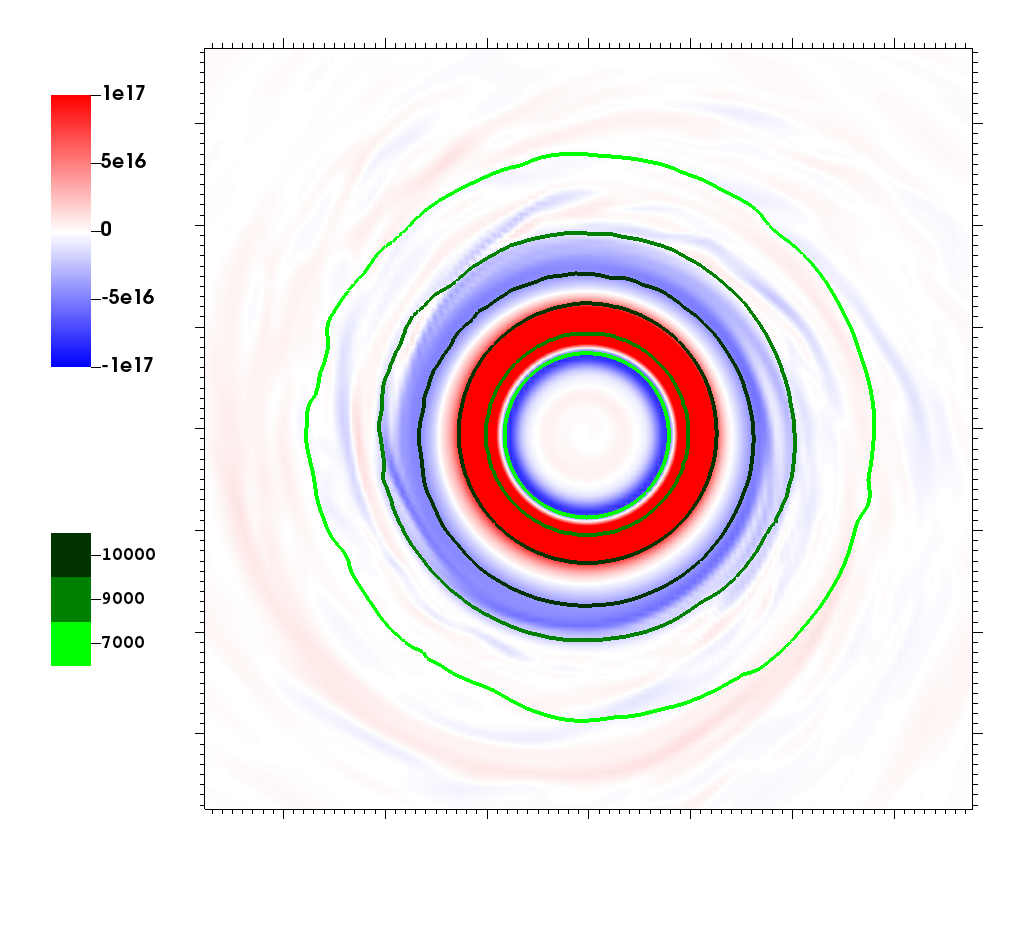} \\
	\includegraphics[width=0.59\linewidth]{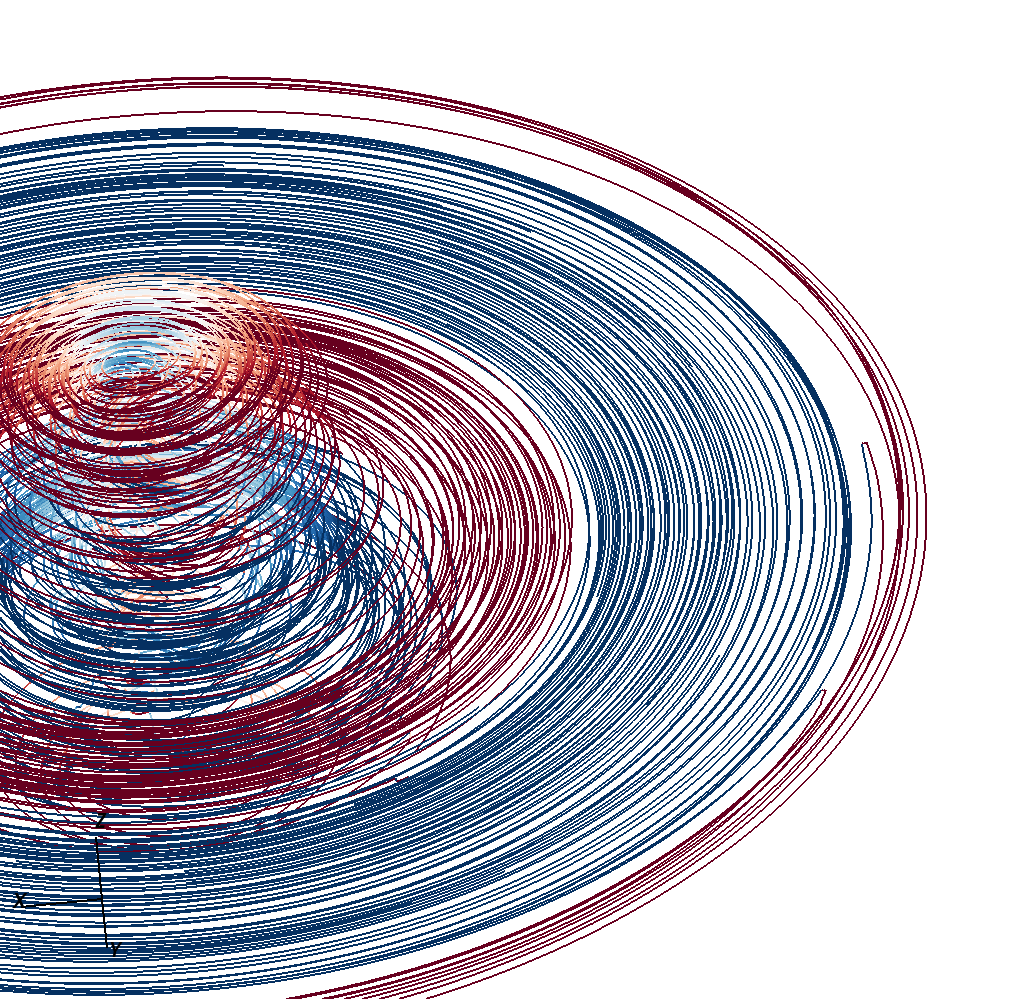}
	\caption{\textit{Magnetic field evolution}. (Top) Toroidal magnetic field intensity in the orbital plane at $t=110~\rm{ms}$ after the merger. The solid colored lines correspond to constant angular velocity surfaces $\Omega=(7000,9000,10000)~\rm{rad/s}$. The toroidal field is large and positive in the region where the angular velocity increases, it is close to zero near the velocity peak and becomes large and negative when then angular velocity decreases.
	(Bottom) Streamlines of the magnetic field. The colors indicate the intensity of the toroidal magnetic field, as in the top panel.}
	\label{fig:b_btor_omega}
\end{figure}

Finally, the above-mentioned magnetic intensity rings seen in Fig.~\ref{fig:slices_B2} can be observed more clearly in Fig.~\ref{fig:b_btor_omega}. It displays the toroidal component of the magnetic field together with iso-surfaces of angular velocity at $t = 110$ ms after the merger. The differential rotation in the remnant amplifies the toroidal component, which changes from clockwise to counter-clockwise depending on radii. The maximum of the angular velocity (where its radial derivative vanishes), matches with a ring of very low magnetic field, as one would expect if such structures were produced by magnetic winding.



\subsection{Energetics evolution}

\begin{figure}
	\centering
	\includegraphics[width=0.9\linewidth]{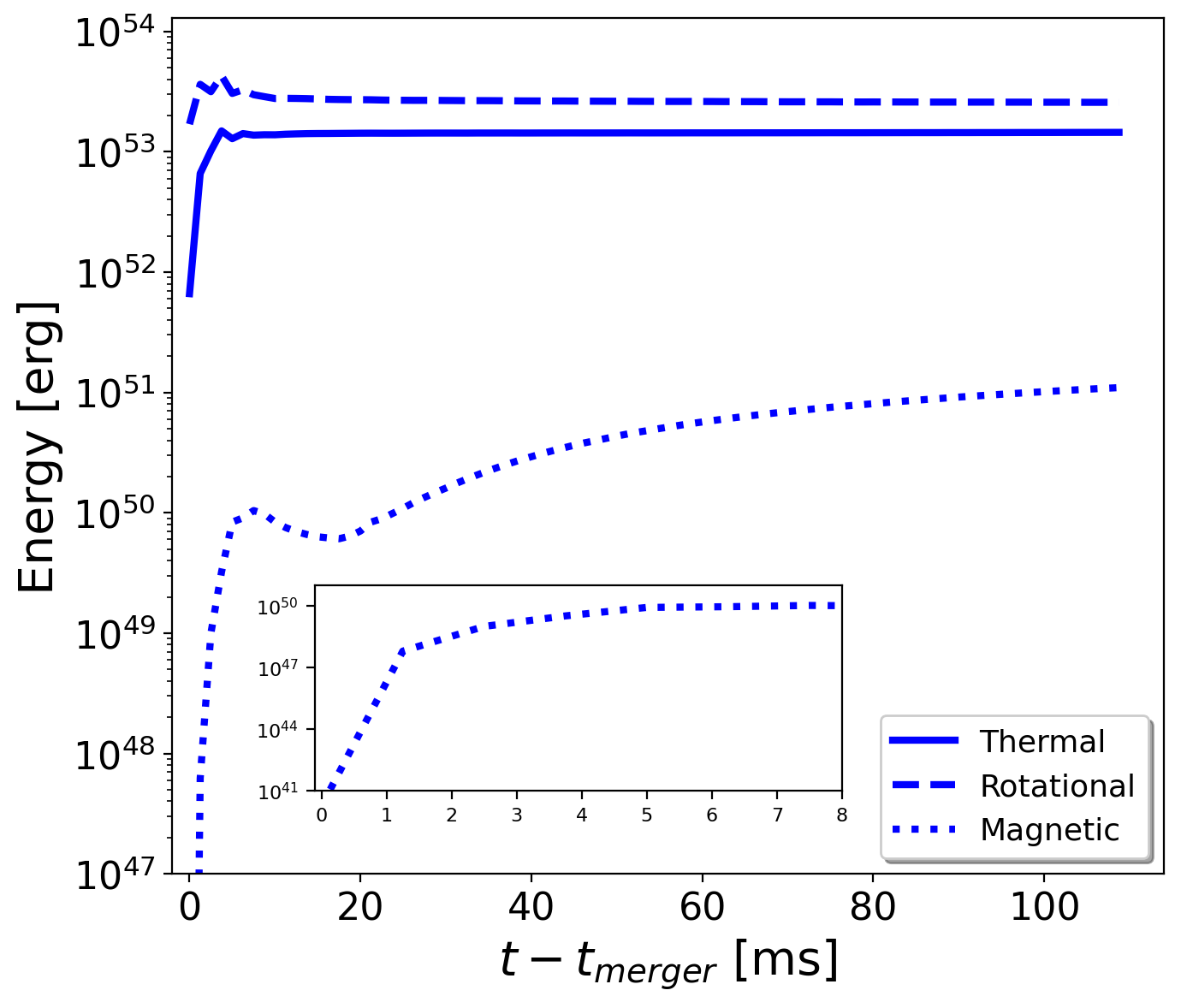}
	\caption{\textit{Energy evolution}. Rotational (dashed), thermal (solid) and magnetic (dotted) energies, integrated over the whole simulation domain.}
	\label{fig:integrals_full}
\end{figure}

\begin{figure}
	\centering
	\includegraphics[width=0.9\linewidth]{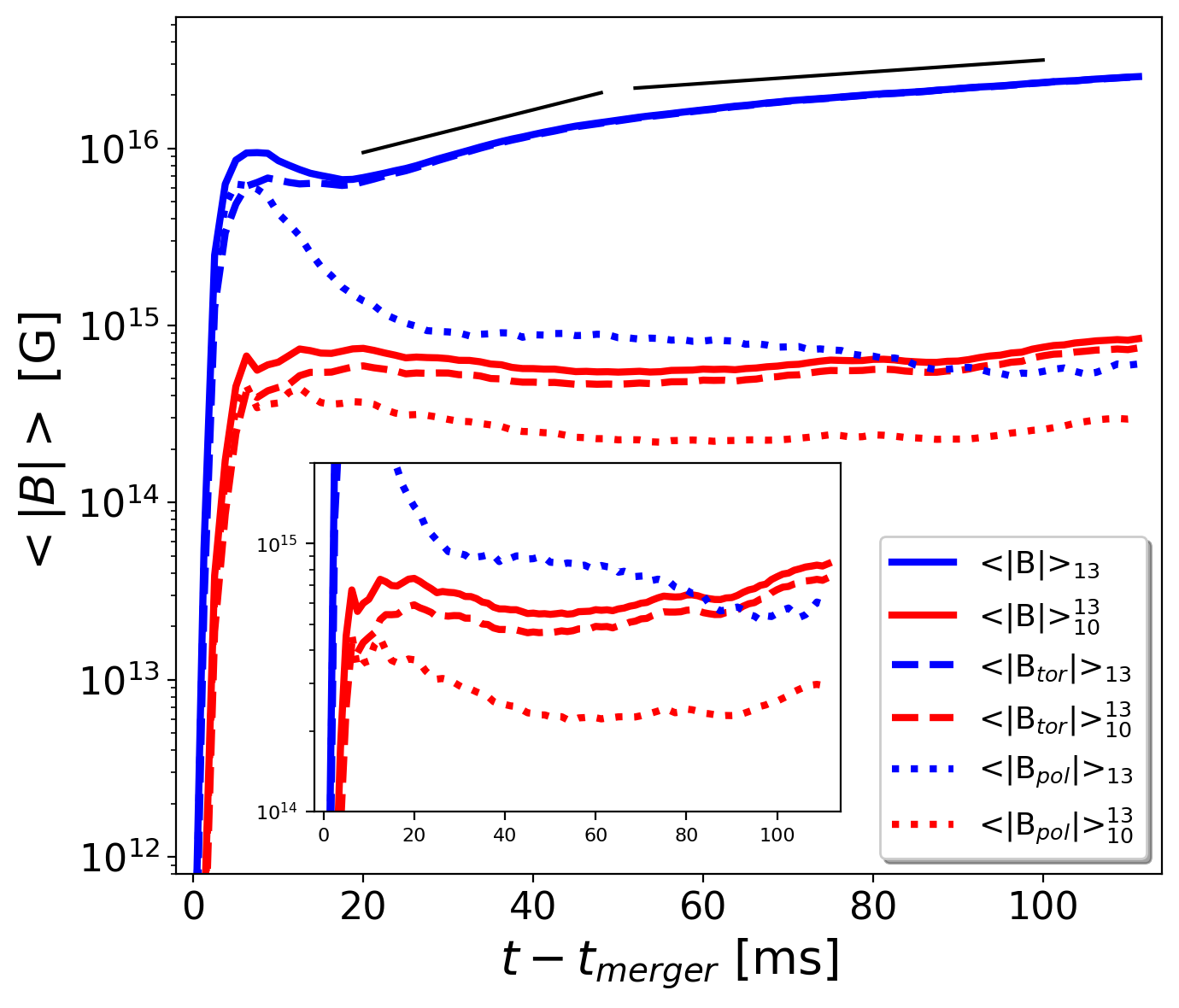}
	\caption{\textit{Average intensity of magnetic field components}. Evolution of the averaged intensity  of the magnetic field (solid), the poloidal (dotted) and toroidal (dashed) components in the bulk and envelope regimes of the remnant. Notice the monotonic growth of the toroidal component for $t \gtrsim 50$~ms, and the increase in the poloidal one for $t \gtrsim 100$~ms. Solid lines correspond to a growth linear in time, as the one expected to be produced by the winding.}
	\label{fig:averagebbpolbtor_full}
\end{figure}

A more quantitative analysis of the results can be achieved by performing global spatial integrations of different relevant quantities. In Fig.~\ref{fig:integrals_full} are displayed the volume-integrated evolution of the thermal, rotational and magnetic energies as a function of time. Few milliseconds after the merger, both thermal and rotational energies have a nearly constant value near $10^{53}$ erg. The magnetic energy is amplified to approximately $10^{50}$ erg in the first $\sim 5$~ms after the merger, and then it rises up to almost $10^{51}$ erg by the end of our simulation at $t \sim 110$~ms.

The growth of the average magnetic field is observed more clearly in Fig.~\ref{fig:averagebbpolbtor_full}. The average magnetic field is decomposed into its toroidal and poloidal components, and integrated either in the bulk or in the envelope. For both regions, the firsts $5$ ms after the merger are dominated by the KHI, responsible of the high amplification of the magnetic field, achieving root mean square values of $10^{16}$~G in the bulk and $10^{15}$~G in the envelope. At later times, the behavior of the magnetic field differs depending on the region. In the bulk, the toroidal part becomes the dominant contribution to the magnetic field. At $\sim 15$~ms after merger, the toroidal component increase linearly in time due to the winding mechanism. The poloidal component, however, decays by one order of magnitude, becoming nearly constant until $\sim 90$~ms after merger.
On the other hand, the envelope shows a constant toroidal magnetic field strength close to $10^{15}$ G and a poloidal component smaller by a factor of a few. At later times, $ \gtrsim 50$~ms, the toroidal component also grows within this region, although it presents some oscillations. This is more clear in the inset of Fig.~\ref{fig:averagebbpolbtor_full}.

\begin{figure*}
	\centering
	\includegraphics[width=0.365\linewidth, trim={0 2.75cm 0 0}, clip]{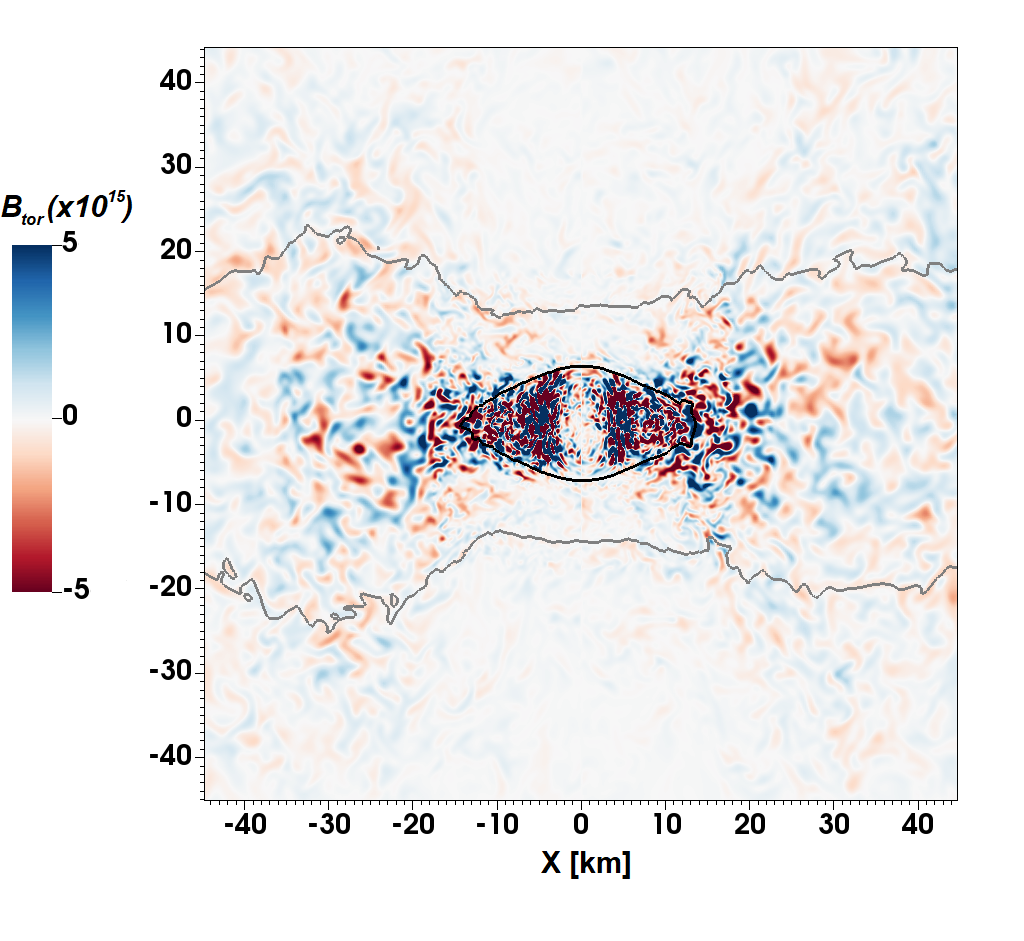}
	\includegraphics[width=0.2975\linewidth, trim={4cm 2.75cm 0 0}, clip]{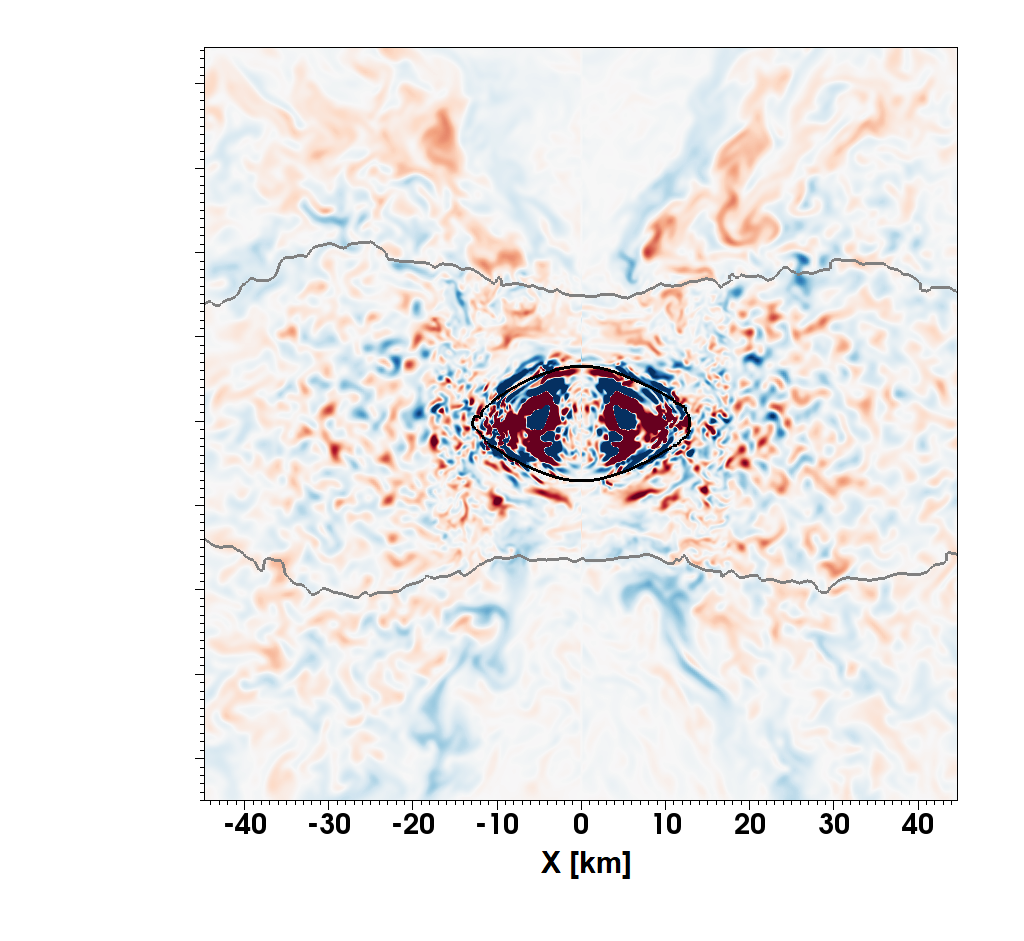}
	\includegraphics[width=0.2975\linewidth, trim={4cm 2.75cm 0 0}, clip]{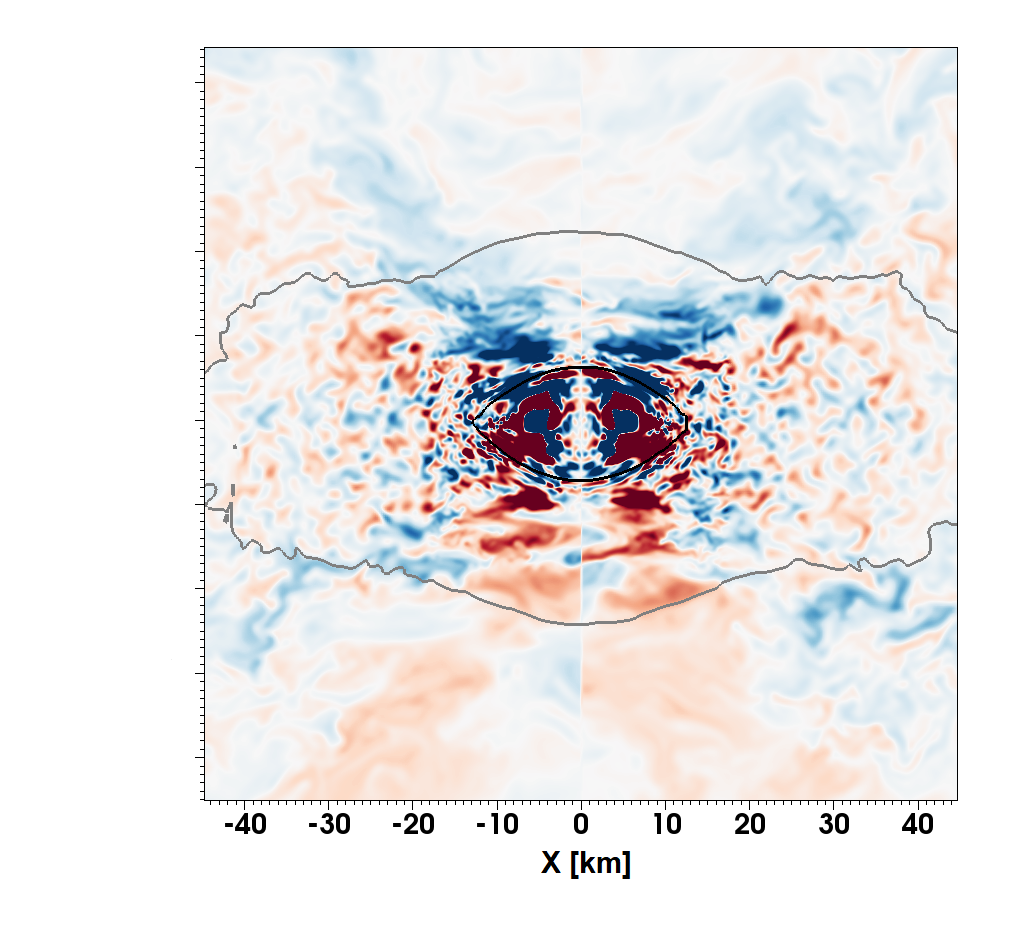} 
	\includegraphics[width=0.365\linewidth, trim={0 1cm 0 0}, clip]{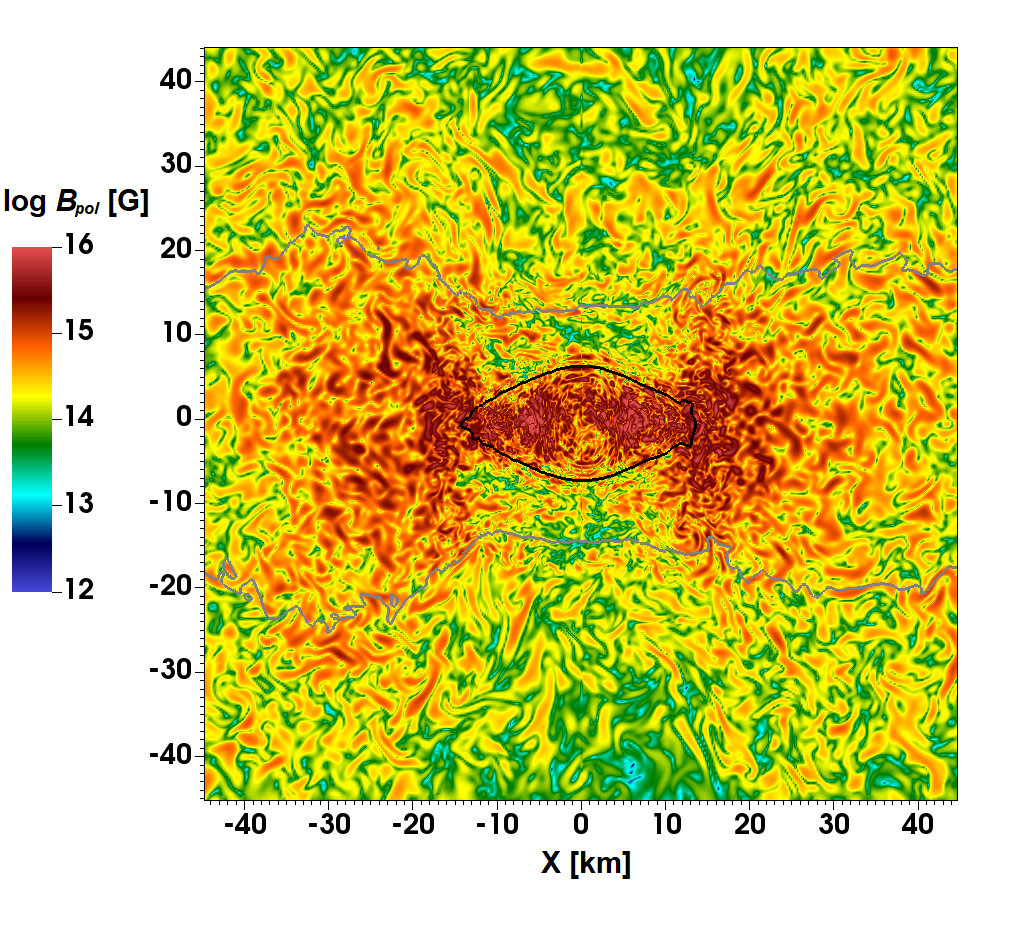}
	\includegraphics[width=0.2975\linewidth, trim={4cm 1cm 0 0}, clip]{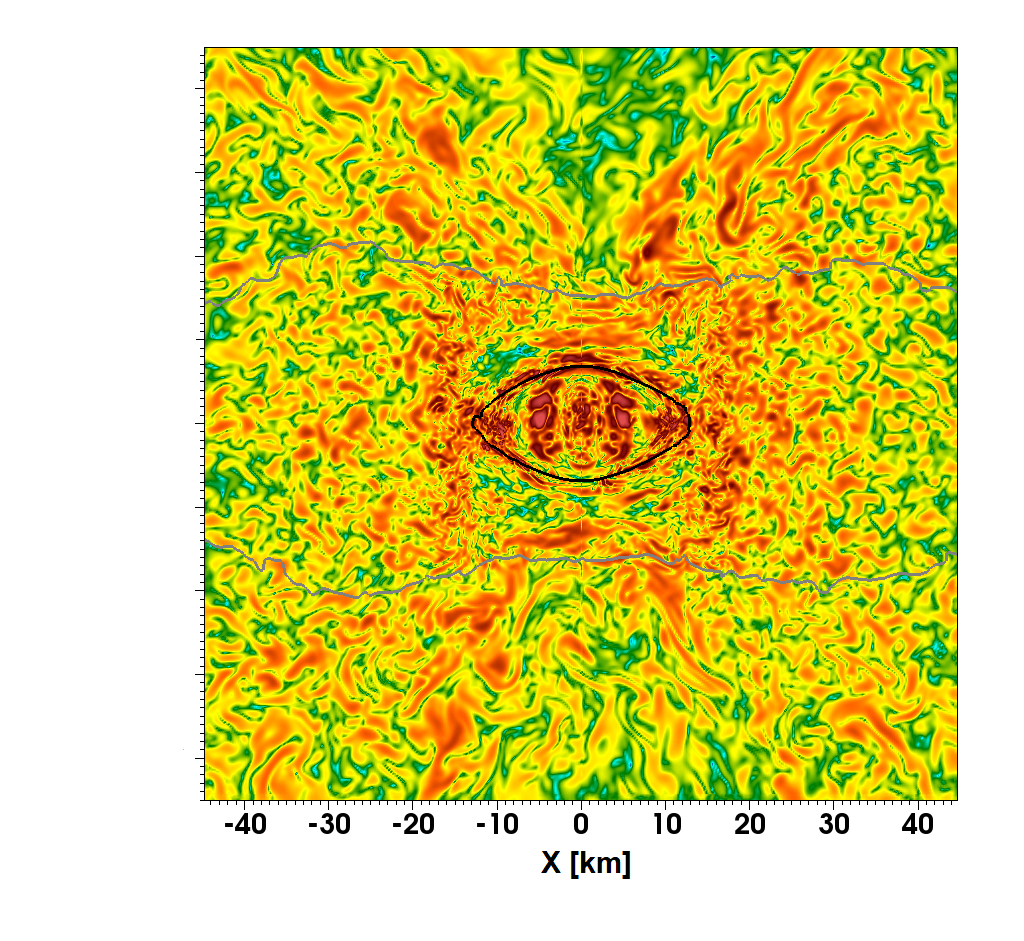}
	\includegraphics[width=0.2975\linewidth, trim={4cm 1cm 0 0}, clip]{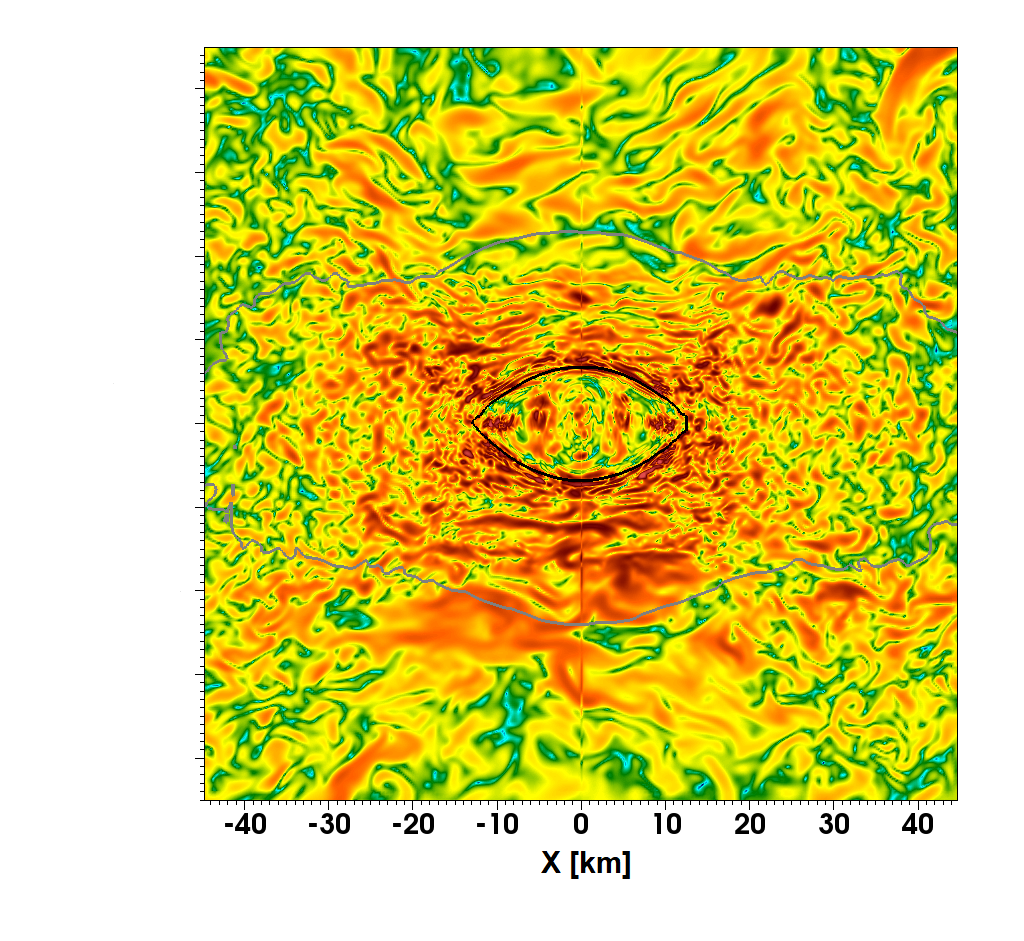} 
	\caption{\textit{Magnetic field components in the meridional plane}. Toroidal (top row) and norm of the poloidal (bottom row) components of the magnetic field at times $t = \{15, 50, 110\}$~ms after the merger. Persistent large-field structures, with a "teddy bear face" shape, are being formed in the bulk for the toroidal component.}
	\label{fig:meridional_Btorpol}
\end{figure*}

\subsection{Spatial distribution}

Some representative time snapshots of the toroidal and poloidal magnetic field components are displayed in Fig.~\ref{fig:meridional_Btorpol}.
Both components are dominated by the turbulent dynamics at the initial stage. However, at later times, persistent intermediate-scale structures of kilometer size develop mostly in the remnant's bulk, first in the toroidal component (with a "teddy bear face" shape) and posteriorly in the poloidal one. Such complex structures, initially produced in a highly turbulent environment, crystallize and remain almost unaltered while expanding and increasing their strength for the rest of the evolution.

\begin{figure}
\centering
	\includegraphics[width=0.9\linewidth]{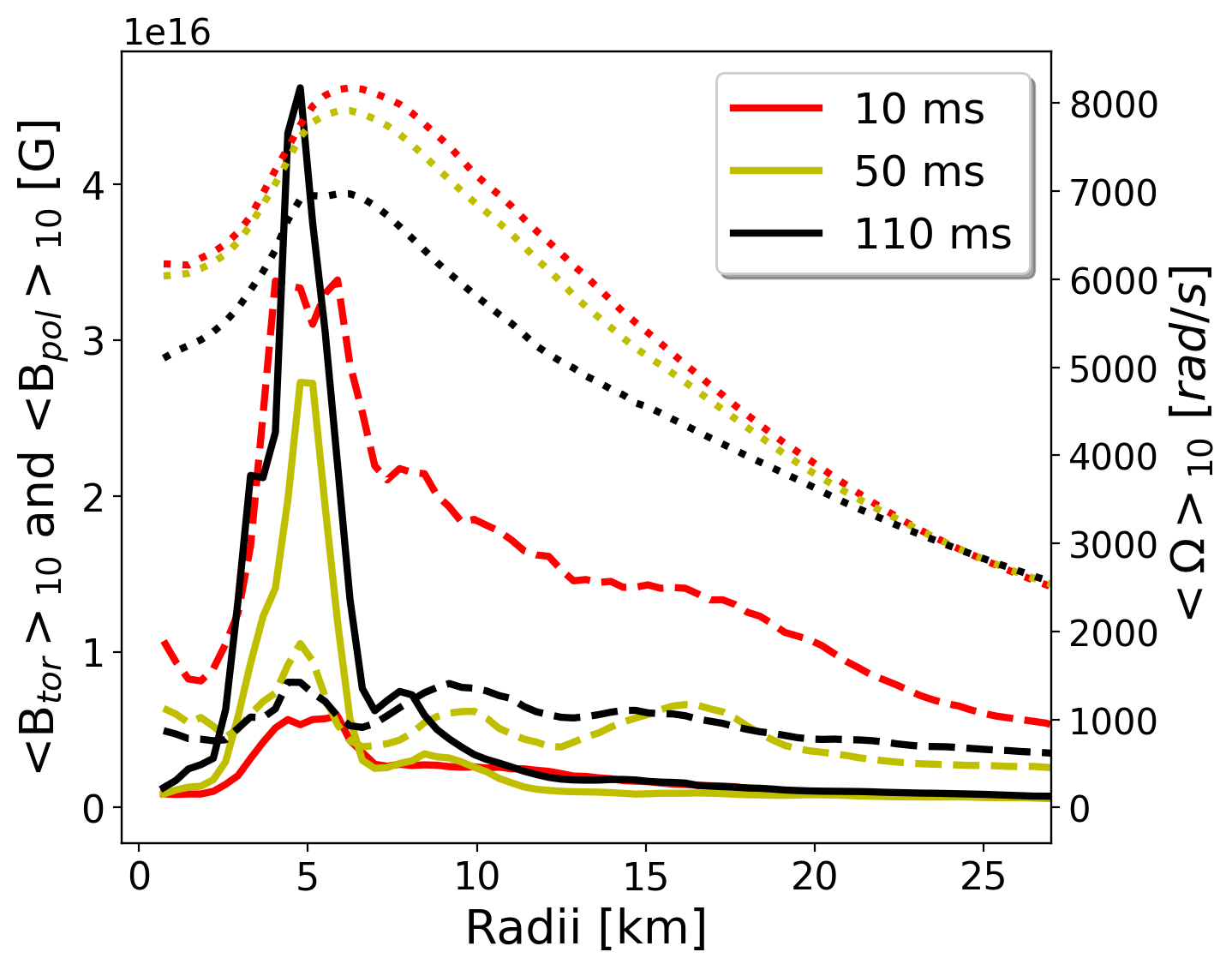}
	\caption{\textit{Envelope+bulk average quantities as a function of the cylindrical radii}. Toroidal (solid lines) and poloidal ($\times 10$, dashed lines) components of the magnetic field. The angular velocity (dotted lines) is represented in the right axis. The poloidal component has been magnified one order of magnitude for visualization purposes. }
\label{fig:cylinder_btor_bpol_omega_full}
\end{figure}

Notice that the distribution of the magnetic field is correlated with the angular velocity  profile. This is shown more clearly in Fig.~\ref{fig:cylinder_btor_bpol_omega_full}, where these fields are plotted as a function of radii for different times after the merger. The poloidal component of the magnetic field (amplified by one order of magnitude in order to see its behavior) reaches maximum values at about $5$~ms after merger. Then, the angular velocity induces an energy transfer from the poloidal to the toroidal component. This is reflected as a decay of the poloidal component at later times. On the other hand, the toroidal component grows linearly due to the winding mechanism at times $\gtrsim 20$ms in the region of the bulk with the largest radial gradient of the angular velocity, approximately located at $\sim 5$ km.

\begin{figure}
\centering
	\includegraphics[width=0.9\linewidth]{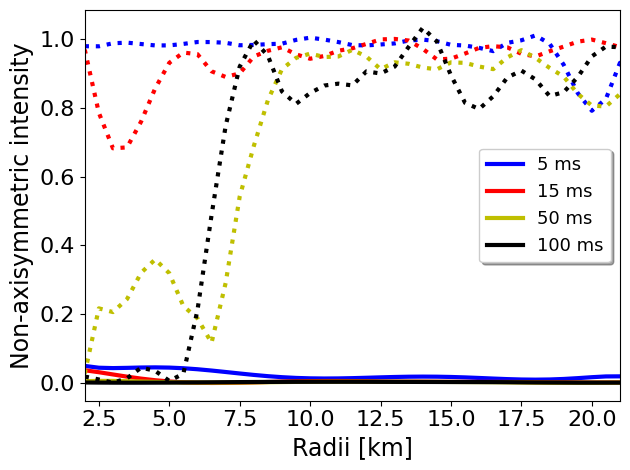}
	\caption{\textit{Estimations of the level of turbulence}. Kinetic (solid lines) and magnetic (dotted lines) non-axisymmetric intensities in the equatorial plane, as defined by Eq.~\eqref{eqn:Iturbulence}. Clearly, the kinetic energy is dominated by the azimuthal average of the velocity field for times $t\gtrsim15$ ms. The magnetic field, on the other hand, is highly non-axisymmetric everywhere at times $t\sim15$~ms. However, at later times, large-scale magnetic field develop in the bulk of the remnant such that it becomes more axisymmetric, indicating that the degree of turbulence drops in that region.}
\label{fig:turbulent_intensity}
\end{figure}

In Fig.~\ref{fig:turbulent_intensity} we have estimated the degree of turbulence of the velocity and magnetic fields in the equatorial plane with the non-axisymmetric indicator definitions given by Eq.~\eqref{eqn:Iturbulence}. At times $t\gtrsim15$~ms the velocity field is highly axisymmetric ($I_{kin} \sim 0.1\%$). On the contrary, the magnetic field is highly non-axisymmetric everywhere at $\sim15$~ms and decays only within the bulk of the remnant at later times, while it is sustained in the envelope at all times. Notice that although the kinetic turbulent energy is relatively small, in absolute values it is still of the same order than the total magnetic energy in this scenario. A visual inspection of the orbital plane in Fig.~\ref{fig:slices_B2} indicates that the residuals responsible for the high values of $I_{mag}$ at $R>10$~km are indeed given by turbulent small scales, although the presence of spiral arm-like intermediate scales is also apparent, especially at the late times (bottom-right panel).

\subsection{Spectral distribution}

\begin{figure*}
	\centering
	\includegraphics[width=\linewidth]{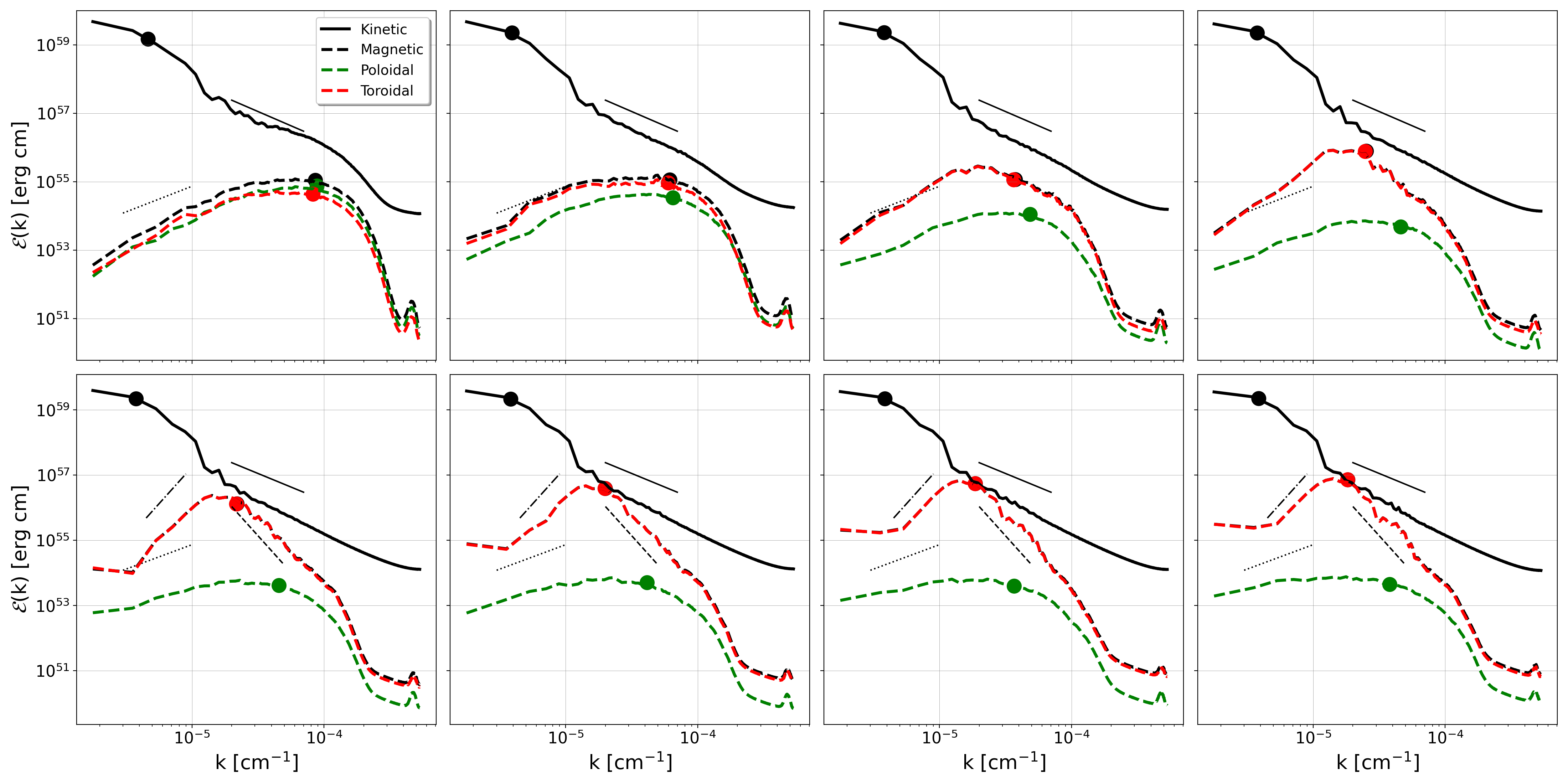}
	\caption{\textit{Magnetic energy spectra}. Kinetic (solid) and magnetic energy spectra (black dashed) as a function of the wavenumber. The toroidal and poloidal components of the magnetic field are represented in red and green colors. From left to right and from top to bottom, $t = \{5,10,20,30,50,80,100,110\}$ ms after the merger. Dotted and solid straight lines correspond to Kazantsev and Kolmogorov slopes, $k^{3/2}$ and $k^{-5/3}$ respectively. The dashed and dashed-dotted  slopes $k^{\pm4.5}$ indicate how the toroidal magnetic energy behaves near the equipartition peak at $k \sim 2 \times 10^{-5}$.}
	\label{fig:spectra_full}
\end{figure*}

\begin{figure}
\centering
	\includegraphics[width=1\linewidth]{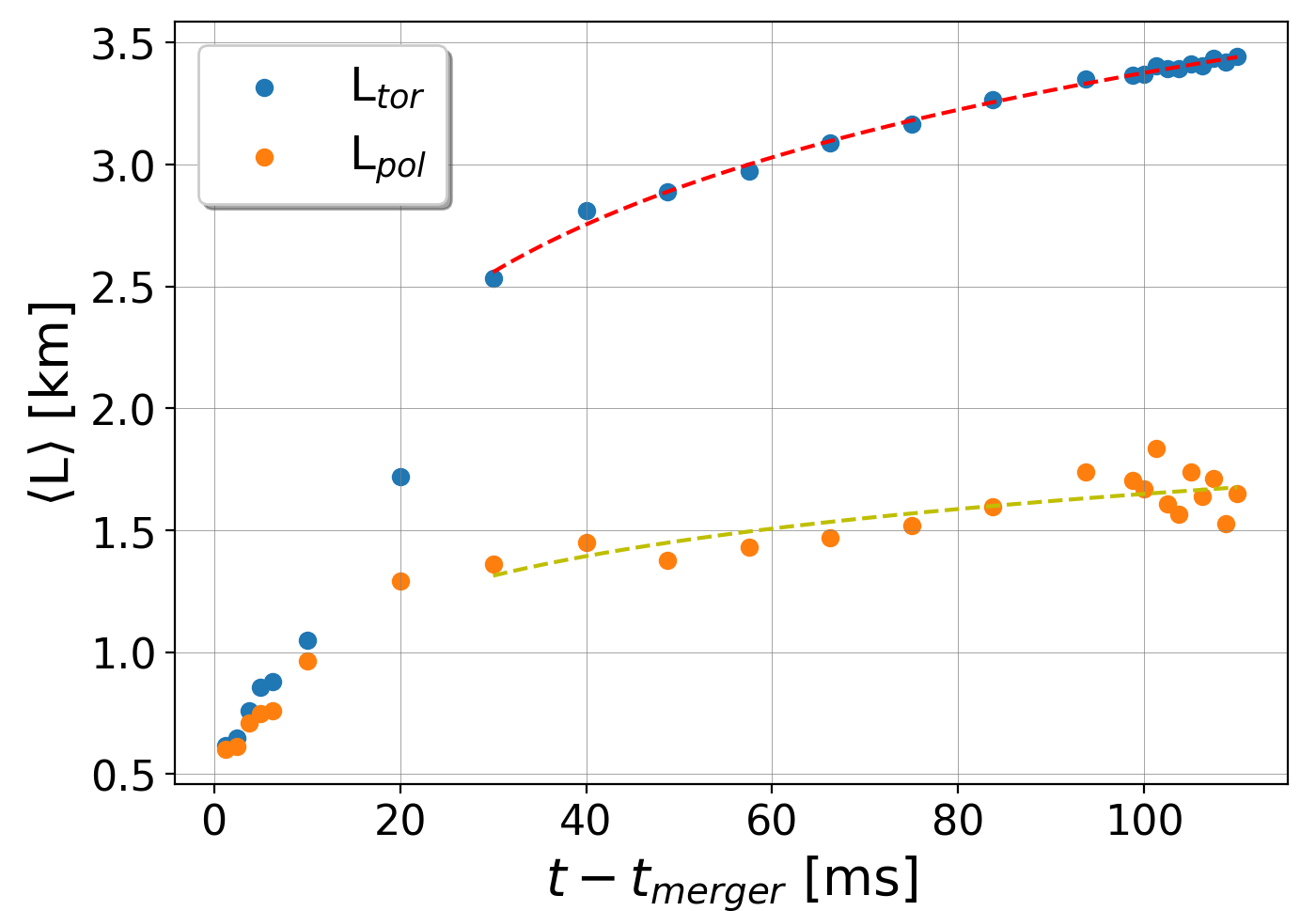}
	\caption{\textit{Average characteristic wavelengths}. 
	Average characteristic length of toroidal (blue) and poloidal (orange) as a function of time. The adjustment corresponds to logarithm curve $\langle L\rangle = a \ \ln(t) + b$ for $t \gtrsim 30$~ms. For the toroidal component, the curve adjusts remarkably well with an equation of $\langle L\rangle = 0.679 \ln(t) + 0.251$~km, while that the poloidal component is well fitted with $\langle L\rangle = 0.279 \ln(t) + 0.366$~km.}
\label{fig:Average_characteristic_wavelength}
\end{figure}

In Fig.~\ref{fig:spectra_full} we compute the kinetic and magnetic energy spectra at different times after the merger $t = \{5,10,20,30,50,80,100,110\}~\rm{ms}$. The solid slope is the Kolmogorov spectra power law and the dotted slope is the Kazantsev one. The solid dots correspond to the averaged wavenumber, which can be used to define a characteristic length of the system. At $t = 5$~ms, the two components of the magnetic energy are almost identical, indicating that they were amplified equally during the isotropic turbulent regime. However, already at $20$~ms after merger, the differences between them are notorious. The toroidal component of the magnetic field changes from Kazantsev to Kolmogorov power law in the intermediate scales, as pointed out in~\cite{palenzuela22}. At later times, a peak in the toroidal component has formed roughly at spatial length of $\sim 3.5$ km. The magnetic energy reaches equipartition with the kinetic one at these length scales at $t \sim 50$ ms. The toroidal component largely dominates during this growing phase, suggesting winding as the main responsible mechanism. For late times, we note two quite symmetric slopes in a narrow intermediate $k$-range, which are best fitted by power laws $\sim k^{\pm 9/2}$, respectively, indicated in the figure. On the other hand, the poloidal component of the magnetic field decays at $20$ ms after merger and gradually flattens.

In Fig.~\ref{fig:Average_characteristic_wavelength}, we have plotted the characteristic length of the toroidal and poloidal magnetic field components as a function of time. 
Soon after the merger the fields are turbulent and the characteristic or coherent scale is very small. However, the turbulent diffusivity and the differential rotation produce a quick growth of the characteristic scales for both components.
For times $t \gtrsim 30$~ms, they adjust remarkably well 
with a logarithmic curve. In particular, by the end of the simulation the toroidal component reaches values $\langle L\rangle \sim 3.5$~km. This is also in agreement with the appearance of spiral-like coherent structures at intermediate scales mentioned before.

\section{Discussion on the magnetic field dynamics} \label{sec:discussion}

Here we will discuss the processes that affect the magnetic field dynamics, together with the evidences from our simulations that supports the emerging picture.

\begin{itemize}
	\item At the merger, a shear layer between the two stars induce a Kelvin-Helmholtz instability. The non-linear interaction of the eddies forming near the shear layer produces a turbulent regime, which expands to all the remnant due to the bouncing of the cores and the rotation (see first and second row of Fig.~\ref{fig:slices_B2}). The magnetic fields are twisted by the turbulent flow and grow exponentially, reaching a \textit{turbulent, quasi-isotropic state with average intensities of $10^{16}~\rm{G}$ in less than $5~\rm{ms}$ after the merger}. There are strong compelling arguments that support these quantitative results. First, the energy spectra at the end of this phase (i.e., $t=5$~ms) follows the expected distributions for a turbulent flow in the kinematic regime, when the magnetic field is still dynamically unimportant (see the top left panel in Fig.~\ref{fig:spectra_full}). In the inertial range, the kinetic spectra decays with the Kolmogorov power-law, whereas the magnetic spectrum at low wavenumbers grows with the Kazantsev power law. The toroidal and poloidal contributions to the magnetic spectra are comparable, as one would expect in an isotropic turbulent regime.
	Besides the consistency of the spectra, there are other evidences: the most relevant is the convergence of our results, in simulations with and without the gradient SGS model for different resolutions~\cite{palenzuela22} and initial magnetic field configurations~\cite{aguilera22}, strongly indicating that the saturation values of the magnetic field and the isotropic turbulent character of the fields at this stage are indeed correct.
	
	\item At $t \sim 5$~ms, turbulence on the magnetic fields is very intense, while that the velocity fields are mostly axisymmetric with small turbulent deviations. This statement is supported by the non-axisymmetric intensities displayed in Fig.~\ref{fig:turbulent_intensity}, which are a proxy to the degree of turbulence. This configuration have two effects. First, the angular velocity breaks the isotropic turbulent state reached during the KHI and will transfer part of the poloidal to toroidal magnetic field in a short timescale $t \lesssim 30$~ms (see in Fig.~\ref{fig:averagebbpolbtor_full} the quick decay at those times of the poloidal component, especially in the bulk).
	Second, in approximately the same timescales, the intense magnetic field turbulence enhances an effective resistivity which leads to the diffusion of the small-scale magnetic fields, transferring energy into larger spatial scales. The magnetic fields reach a coherent or characteristic length scale $\sim 2.5$~km at $t\sim30$~ms, as it can be observed in Fig.~\ref{fig:Average_characteristic_wavelength}. 
	
	\item At this point, when the coherent scales are sufficiently large, the winding mechanism becomes efficient and amplifies linearly the toroidal component by shearing the poloidal one, as it is shown in Fig.~\ref{fig:averagebbpolbtor_full}. 
	The effective turbulent resistivity decreases in the bulk region as the magnetic turbulence becomes weaker (see the monotonic decay for $t \gtrsim 15$~ms in  Fig.~\ref{fig:turbulent_intensity}). This is consistent with the evolution of the characteristic magnetic field scale displayed in Fig.~\ref{fig:Average_characteristic_wavelength}, where in the early times the small scales govern the dynamics, but it rapidly restructures into larger scales in approximately the first $30$ ms after the merger. Later on, the growth of the  coherent magnetic field scale is slowed down.
	
	\item As the magnetic field acquires a more and more ordered large-scale toroidal structure, the fluid can be potentially subject to the MRI in the region where $\partial_R \Omega < 0$. The angular velocity at $R\gtrsim 6-7$ km follows a nearly-Keplerian profile $\Omega \sim R^{-3/2}$, satisfying the previous condition. The trends seen in Fig.~\ref{fig:turbulent_intensity}, where $I_{mag}\sim 1$ only in the region where MRI is allowed, are possibly compatible with this scenario. A more definitive proof would follow from the evaluation and comparison of the simulated and predicted excited wavelengths $\lambda_{MRI}$, a hardly possible task in this scenario given the very complex background and residual topology (i.e., the classical and widespread used evaluations of $\lambda_{MRI}$ assume a background, stable, homogeneous field, totally incompatible with the one observed here). In particular, the observed formation of spiral-like, coherent intermediate scales at very late time, contributing to $I_{mag}$, needs to be understood more in detail. Regardless on the formal labeling of the process, the non-linear mechanism acting in the differentially rotating envelope sustain a highly complex magnetic fields, composed of small and intermediate scales, for all times spanned in our simulation. This same mechanism begins to convert toroidal into poloidal field for $t \gtrsim 90$~ms (see again Fig.~\ref{fig:averagebbpolbtor_full}).
\end{itemize}

For a more quantitative modeling to explain the previously discussed results, let us consider an almost axi-symmetric configurations in cylindrical coordinates $(R,\phi,z)$. Velocity and magnetic fields can be decomposed into poloidal and toroidal components
\begin{equation}
	{\bf v} = {\bf v_P} + \Omega R {\hat \phi}
	~~,~~
	{\bf B} = {\bf B_P} + B_{\phi} {\hat \phi}
\end{equation}
where $\Omega = \Omega(R,z)$ and ${\bf B_{P}} = \nabla \times {\bf A} = \nabla \times (A {\hat \phi})$. Within these assumptions, it is straightforward to write down the induction equation as (see e.g. Section 6.3 in \cite{mestel2003stellar})
\begin{eqnarray}\label{B_winding_mestel1}
	\partial_t A + \frac{1}{R}{\bf v_P} \cdot \nabla (R A) &=& \alpha B_{\phi}  \\
	&+& \eta\, (\nabla^2 - R^{-2}) A \nonumber \\ 
	\label{B_winding_mestel2}
	\partial_t B_{\phi} + R \nabla \cdot \left( \frac{B_{\phi} {\bf v_P}}{R} \right) &=& R\, {\bf B_P} \cdot \nabla  \Omega 
	\\
	&+& \eta (\nabla^2 - R^{-2}) B_{\phi} \nonumber~~.
\end{eqnarray}
These equations describes mainly the advection of both components with the poloidal velocity, modified by source terms which account for three different effects.

The first term in the right-hand-side of Eq.~\eqref{B_winding_mestel2} represents the familiar winding mechanism, i.e., the generation of toroidal field by the shearing of the poloidal one due to differential rotation. 
The other two terms, proportional to $\alpha$ and $\eta$, appear only when decomposing the induction equation into a large-scale mean field and small-scale turbulent residual. With that decomposition, the mean electromotive force can be linearly expanded as a function of the magnetic field, generating these two new terms~\cite{mestel2003stellar}.
The first term in the right-hand-side of Eq.~\eqref{B_winding_mestel1} predicts a growth of the poloidal component, proportional to the toroidal component $B_{\phi}$ and parametrized by the turbulent coefficient $\alpha$.
The last terms in both Eqs.~(\ref{B_winding_mestel1}) and (\ref{B_winding_mestel2}) are diffusion terms proportional to the turbulent resistivity $\eta$. These terms are necessary to diffuse the small scales  and produce large-scale fields.  
Therefore, the large-scale differential rotation generates toroidal from the poloidal magnetic field, while that the small-scale convective eddies are where the toroidal is converted into poloidal. Turbulent resistivity would facilitate the diffusion from small to large scales.

Let us simplify the system of Eqs.~(\ref{B_winding_mestel1})-(\ref{B_winding_mestel2}) by considering very small convective motions ${\bf v_P} \approx 0$. In a nearly quasi-stationary state both the angular velocity and the poloidal magnetic field vary slowly with time $\partial_t {\bf B_P} \approx \partial_t \Omega \approx 0$. Assuming also that the small-scale effects are negligible (i.e., $\alpha \approx \eta \approx 0$), the toroidal component $B^{\phi} = B_{\phi}/R$ evolves as 
\begin{eqnarray}\label{B_winding}
	&&B^{\phi} (t) - B^{\phi} (t_0) =  \left[ B^R \partial_R \Omega + B^z \partial_z \Omega \right] (t - t_0)  \\
	& & \approx  10^{16} G \left( \frac{B^R}{10^{15}\mathrm{G}}\right) \left( \frac{\Delta \Omega}{2 \mathrm{rad/ms}}\right) \left( \frac{4 \mathrm{km}}{\Delta R}\right) \left( \frac{t }{15 \mathrm{ms}}\right) \nonumber
\end{eqnarray}
where we have used typical values for the fields in the bulk of the remnant. This simple estimate shows that the toroidal magnetic field can grow linearly to ultra-strong levels in very short timescales.

Further insight on these simplified solutions can be obtained by considering first the case where the gradients of the angular velocity along the z-direction are much smaller than along the radial direction (i.e., $\partial_R \Omega >> \partial_z \Omega$).
In the remnant produced by the binary merger, the angular velocity is non-monotonic: it grows from its central value (i.e., $\partial_R \Omega >0$), reaches a maximum $\Omega_{max}$ at $R \approx 6$ km, and then it decays at larger radii with a nearly Keplerian profile $\Omega(R) \sim R^{-3/2}$ (i.e., $\partial_R \Omega \approx -3 \Omega/(2 R) <0$). According to Eq.~\eqref{B_winding}, this implies that the toroidal component will remain constant in the peak where $\partial_R \Omega =0$, but it will grow in time with different signs on each side of the peak. This switch in the sign of the toroidal magnetic field component can be clearly observed in Fig.~\ref{fig:b_btor_omega}. 

We can also consider the opposite case, by allowing only gradients of the angular velocity  along the z-direction. Let us consider an angular velocity profile $\Omega(R,z)$ which decreases as we move out of the $z=0$ plane. The shear in the $z$-direction generates torsional Alfvén waves along each field line, propagating away from the equatorial plane. If this shearing is sustained in time, it will form an helicoidal magnetic field structure equivalent to the one presented in Fig.~\ref{fig:3D_streamlines}.

\section{Confirmation of universality of the evolved topology} \label{sec:outermost_layers}

\begin{figure*}
	\centering
	\includegraphics[width=0.32\linewidth, trim={6.5cm 2.5cm 0 0}, clip]{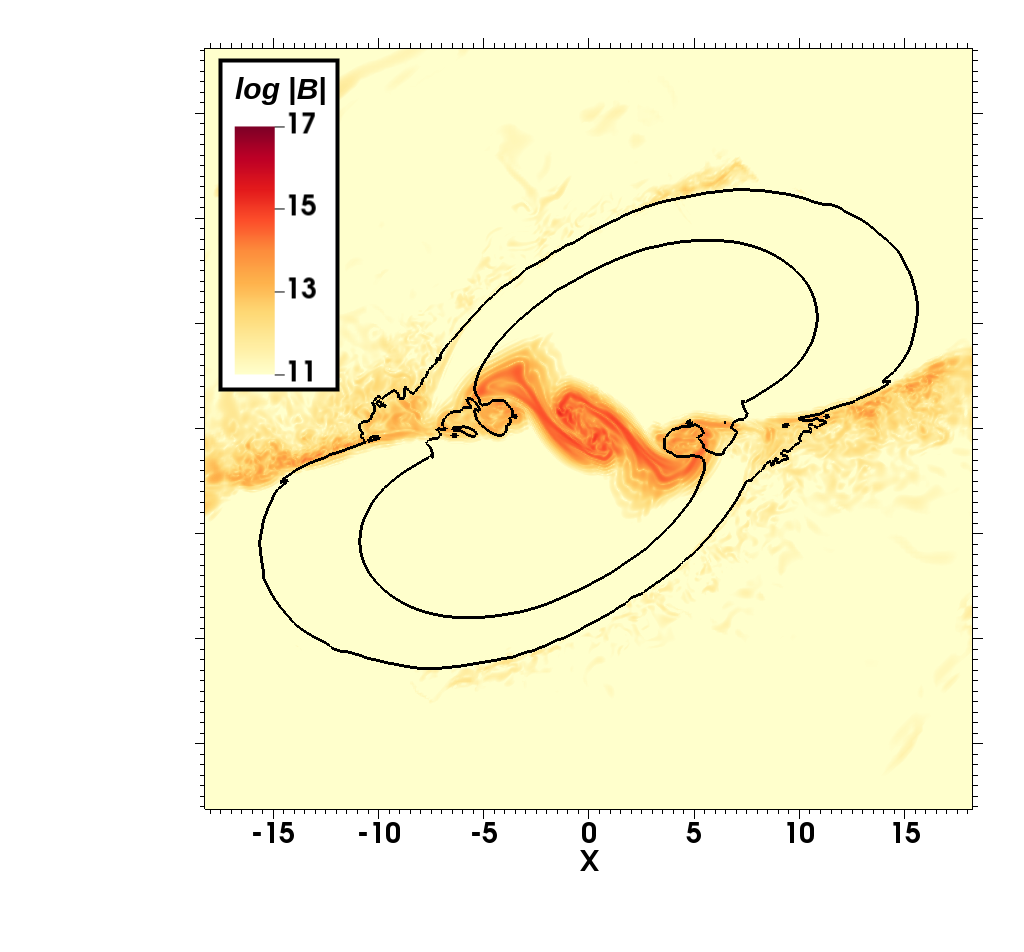}
	\includegraphics[width=0.32\linewidth, trim={6.5cm 2.5cm 0 0}, clip]{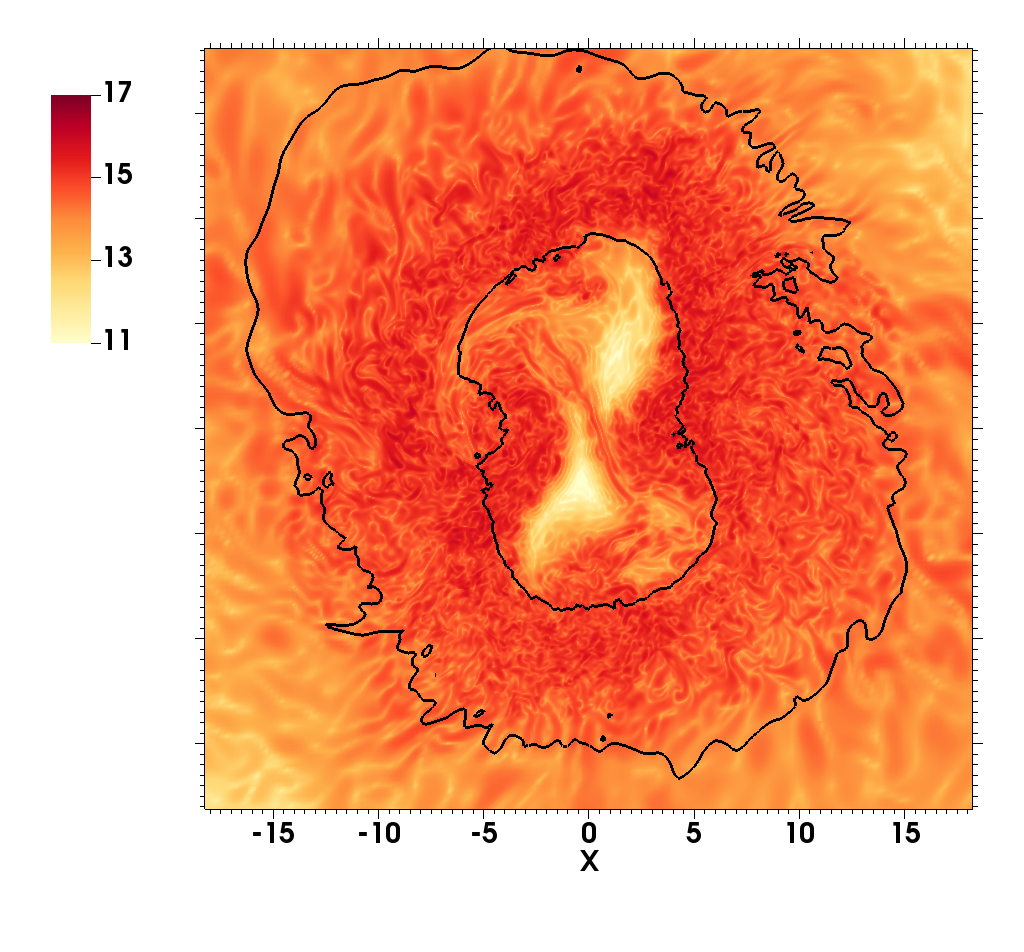}
	\includegraphics[width=0.32\linewidth, trim={6.5cm 2.5cm 0 0}, clip]{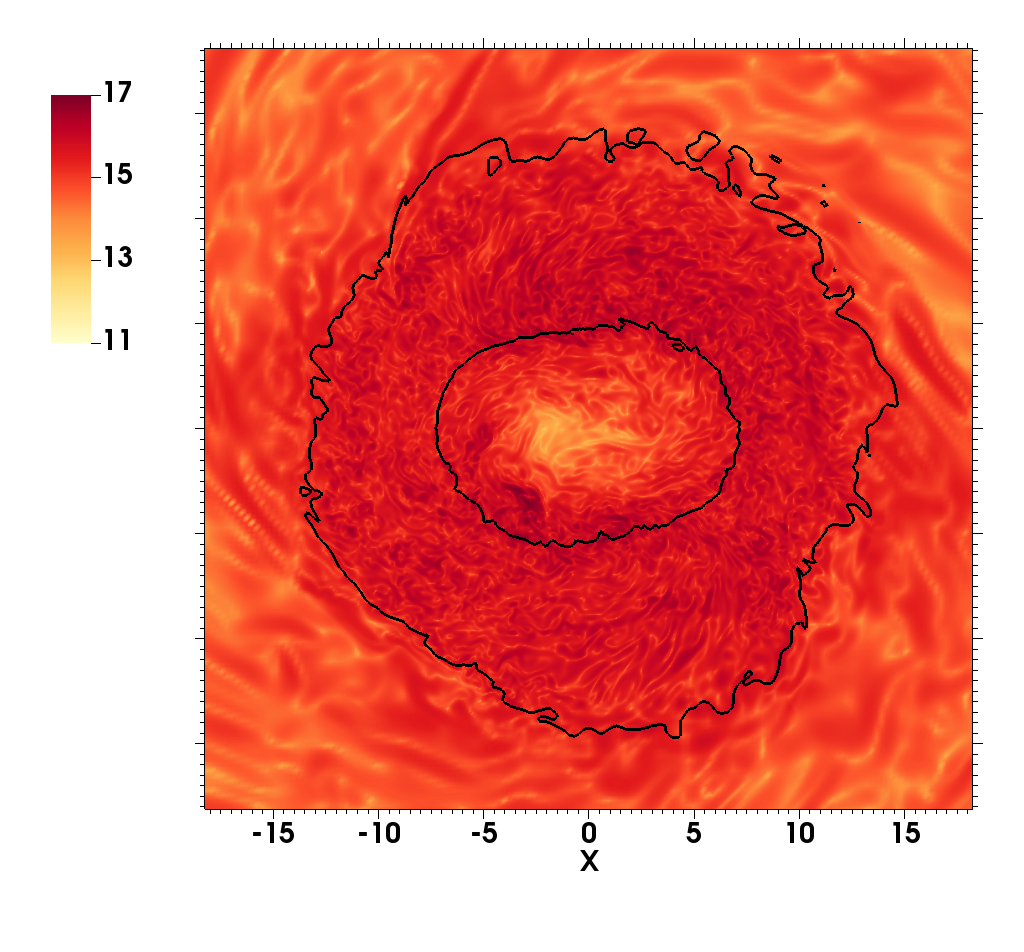}
	\caption{\textit{Magnetic field confined in the crust}. Values of the magnetic field intensity $|\vec{B}|$ in the orbital plane at $t=\{0.5,2.5,5\}$ ms after the merger. Outer and inner black lines correspond to constant density contours ith $\rho = 10^{13}$ and $5 \times 10^{14}~\rm{g~cm^{-3}}$, respectively. The length is given in [km].}
	\label{fig:slices_B2bis}
\end{figure*}

It has been suggested recently that magnetic fields confined in the outermost layers could lead to a different final configuration in the remnant than when they are distributed through all the star~\cite{chabanov23}. Here we also address this question by comparing our long simulation, in which  the magnetic field is initially distributed over all the star, with another one in which the initial magnetic field is confined only to the outermost layers. 
This new simulation is evolved only up to $7.5$ ms after the merger. Following~\cite{chabanov23}, we set the potential vector to $A_i \propto (x^j - x_{NS}^j) \epsilon_{ij} \exp\left[-g_w(r -g_r )^2\right] \max(P - P_{cut}, 0)^n$, being $x_{NS}^j$ the coordinate centers of the two stars and $\epsilon_{ij}$ is the Levi-Civita symbol. We use the same values for the parameters $( g_w,g_r,n )$ as in~\cite{chabanov23}. 


Fig.~\ref{fig:slices_B2bis} shows the evolution of this merger at $t=\{0.5,2.5,5\}$ ms after the merger in the orbital plane, to be compared with Fig.~\ref{fig:slices_B2}. 
As it can be seen from the slices, the magnetic field also is amplified up to local values of $10^{17}$~G in less than $5$~ms after the merger. By comparing both Fig.~\ref{fig:slices_B2} and Fig.~\ref{fig:slices_B2bis}, no significant visual differences can be appreciated in the remnant after those times.

\begin{figure}
	\centering
	\includegraphics[width=0.7\linewidth]{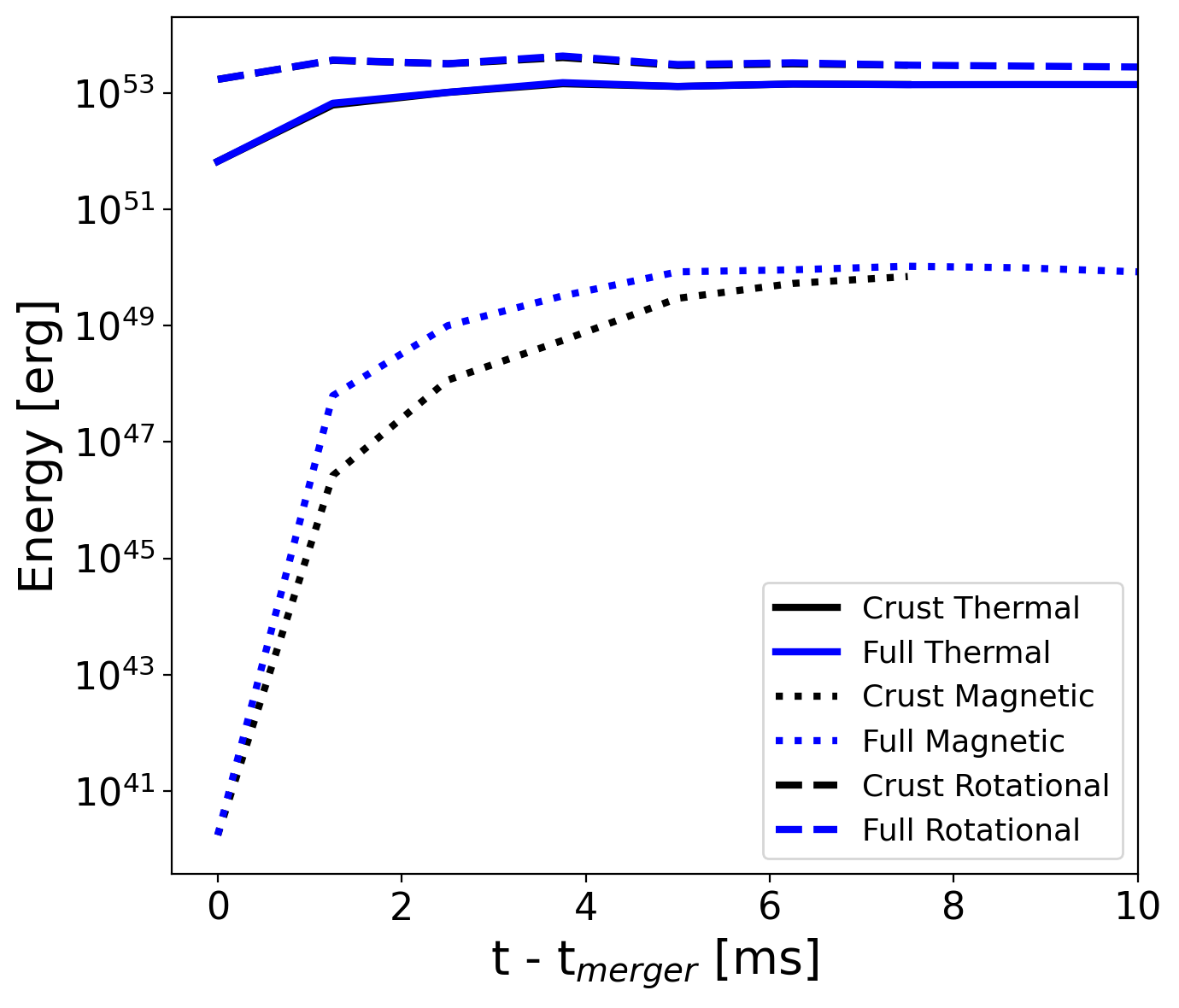}
	\caption{\textit{Energy evolution}. Rotational (dashed lines), thermal (solid lines) and magnetic (dotted) energies, integrated over the whole dominion, for different simulations as a function of time. Notice the comparable values reached by the magnetic energies.}
	\label{fig:integrals}
\end{figure}


In Fig.~\ref{fig:integrals}, we represent the volume-average energies for both simulations (thermal, rotational and magnetic) as a function of time. The results obtained are comparable for the magnetic energies, with negligible differences in both thermal and rotational energies. 

\begin{figure*}
	\centering
	\includegraphics[width=\linewidth]{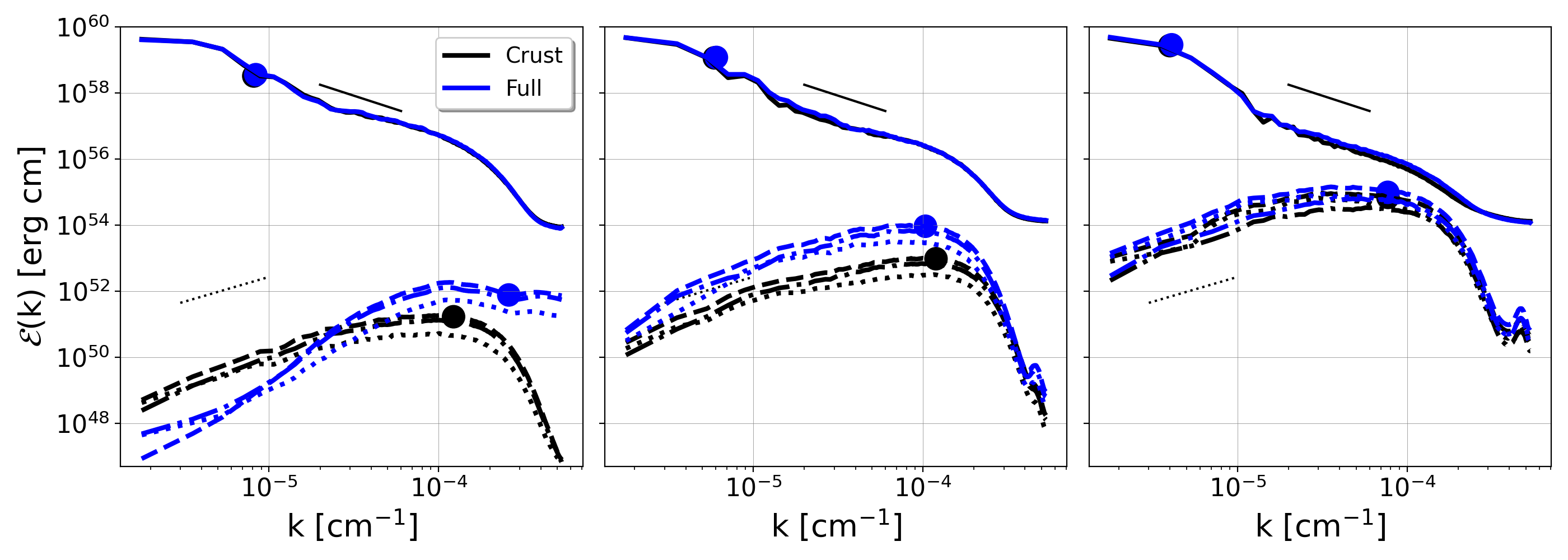}
	\caption{\textit{Energy spectra}. Kinetic (solid), magnetic (dashed) energy spectra for different configurations as a function of the wavenumber at, from left to right, $t=\{1.25,2.5,7.5\}$ ms after the merger. The magnetic one has also being decomposed into its toroidal (dotted) and poloidal (dashed-dotted) components. The solid thin black line corresponds to the Kolmogorov slope, while the dotted one belongs to the Kazantsev slope. The large dots are the wavenumbers which contains, in average, most the energy of each spectra. Although there are some differences at early times, they are negligible at times $t \gtrsim 5$~ms after the merger.}
	\label{fig:spectra_all}
\end{figure*}

Fig.~\ref{fig:spectra_all} shows the spectral energies comparison between the two simulations at different times after merger, i.e. $t=\{1.25,2.5,7.5\}$ ms. 
Although at short times there are some differences on the magnetic energy spectrum, they vanishes at later times.
This suggests that the KHI also amplifies efficiently the magnetic energy in the very first milliseconds after merger, even if they are initially confined exclusively within the outermost layers.

\section{Conclusions} \label{sec:conclusions}

In this paper we have performed LES with the gradient SGS model to accurately capture the magnetic field dynamics during and after BNS mergers. This simulation, although similar to the medium-resolution LES performed in \cite{palenzuela22}, includes two significant differences. First, the SGS gradient model is applied not only in the bulk but also in the envelope, which allows us to include the effects of small-scales into the larger ones also in that region. Second, the simulation is extended up to $110$~ms after the merger, sufficiently long timescale as to study the relevant dynamics occurring in the remnant.

During the merger, there is a turbulent amplification  phase triggered by the KHI. In the bulk region of the remnant, the average magnetic field intensity is exponentially amplified up to $10^{16}$~G by this mechanism in less than $\sim 5$~ms after the merger. Since the turbulence is nearly isotropic, both toroidal and poloidal magnetic field components reach similar values.
From there on, the enhanced turbulent resistivity and then the winding up effect are the main drivers of the magnetic field dynamics. The turbulent resistivity enhances the diffusion of small-scales, developing magnetic fields in coherent scales of a few km in $t \lesssim 30$~ms. The winding mechanism acts at all times. At $t \lesssim 30$~ms, it mainly converts poloidal into toroidal field. Then, when the coherent magnetic field scales are large enough, it becomes more efficient and it is responsible for the linear-in-time growth of the toroidal component of the magnetic field, creating an helicoidal structure. The average intensity of the mainly turbulent poloidal component remains merely constant up to $t \sim 90$ ms, when it starts to grow. Although our results suggest that a magnetically dominated helicoidal structure is slowly being formed, it needs longer times to fully develop under these conditions. The inclusion of neutrino effects might accelerate this process, cleaning the funnel region near the $z$-axis and allowing the magnetic field tower to expand along that direction more easily.

We strongly remark that {\bf our results do not imply that one can effectively model the exponential amplification produced during the KHI by starting with a strong, large-scale, poloidal magnetic field}. In that case, the final state might be already contaminated by the large-scale magnetic field, leading to an accelerated growth due to the winding mechanism and ignoring completely the dominant, small-scale structures. The posterior evolution might be, at best, shifted in time with respect to the correct one. In the worst case, the non-linear dynamics might produce unrealistic results (i.e., like the early production of jets when there should be none, since they are facilitated by large-scale magnetic fields). Therefore, we want to stress that our simulations (present and past works \cite{palenzuela22,aguilera2020,aguilera22}) indicate that {\bf in the absence of enough numerical resolution, the use of strong magnetic fields could be physically acceptable only if their topology is dominated by an axisymmetric toroidal component with highly turbulent homogeneous perturbations, as seen for the saturation state after the KHI phase.}

Finally, we have compared the magnetic field dynamics when the initial magnetic field is either confined in the outermost layers or distributed over all the star. We found that the magnetic field amplification is similar in these two cases. They are also comparable to other simulations presented in \cite{aguilera22}, in which we studied specifically the dependence on the initial configurations of the magnetic field. On the contrary, significant differences have been observed in \cite{chabanov23}. There might be several reasons which could explain such divergent results. One explanation could be the effective numerical resolution. In~\cite{palenzuela22}, the minimum resolution needed in order to obtain convergence of the magnetic field during the KHI was at least $\Delta x =30$~m in simulations with high-order schemes without the inclusion of the gradient SGS model. By using the LES, the resolution requirements were less restrictive, and similar results were obtained with $\Delta x =60$~m. The simulations in~\cite{chabanov23} were performed using similar high-order schemes, but with a maximum resolution of $\Delta x =70$~m and without any SGS model, which might have not been enough to faithfully capture all relevant scales developing in the KHI. Another possible explanation could be that the EoS employed in~\cite{chabanov23} do not allow much bouncing of the cores during the merger, limiting severely the redistribution of the magnetic fields amplified in the shear layer through the rest of the remnant. In particular, when the amplified magnetic fields remain in the envelope without digging its way to the bulk, they might be ejected at later times with the disrupted matter. The possible dependence of the EoS in the final distribution of the magnetic field is an interesting issue currently under study.  
\subsection*{Acknowledgements}

We thank Stephan Rosswog for useful discussions. This work was supported by European Union FEDER funds, the Ministry of Science, Innovation and Universities and the Spanish Agencia Estatal de Investigación grant PID2019-110301GB-I00. RA-M is funded by he Deutsche Forschungsgemeinschaft (DFG, German Research Foundation) under Germany’s Excellence Strategy – EXC 2121 ”Quantum Universe” – 390833306. DV is funded by the European Research Council (ERC) Starting Grant IMAGINE (grant agreement No. [948582]) under the European Union’s Horizon 2020 research and innovation programme. DV's work was also partially supported by the program Unidad de Excelencia María de Maeztu CEX2020-001058-M. The authors thankfully acknowledge the computer resources at MareNostrum and the technical support provided by Barcelona Supercomputing Center (BSC) through Grant project AECT-2022-3-0002 by the RES regular call and  the facilities of the North-German Supercomputing Alliance (HLRN).

\bibliographystyle{unsrt}
\bibliography{turbulence}

\end{document}